\documentclass[showpacs,10pt,aps,prd]{revtex4}
\usepackage{eurosym}
\usepackage{amsfonts}
\usepackage{graphicx}
\usepackage{amsmath}
\usepackage{amssymb}
\usepackage{amsthm}
\def\be{\begin{equation}}
\def\ee{\end{equation}}
\def\bea{\begin{eqnarray}}
\def\eea{\end{eqnarray}}

\newcommand{\mc}[1]{\mathcal{#1}}
\newcommand{\f}[2]{\frac{#1}{#2}}

\begin{document}

\title{The Maxwell-Chern-Simons gravity and its cosmological implications}
\author{Zahra Haghani$^1$}
\email{z.haghani@du.ac.ir}
\author{Tiberiu Harko$^{2,3}$}
\email{t.harko@ucl.ac.uk}
\author{Shahab Shahidi$^1$}
\email{s.shahidi@du.ac.ir}
\affiliation{$^1$ School of Physics, Damghan University, Damghan, Iran,}
\affiliation{$^2$ Department of Physics, Babes-Bolyai University, Kogalniceanu Street,
Cluj-Napoca 400084, Romania,}
\affiliation{$^3$Department of Mathematics, University College London, Gower Street,
London WC1E 6BT, United Kingdom.}

\begin{abstract}
	We consider the cosmological implications of a gravitational theory containing two vector fields coupled minimally to gravity as well as a generalized Chern-Simons term that couples the two vector fields.  One of the vector fields is the usual Maxwell field, while the other is a constrained vector field with constant norm included in the action via a Lagrange multiplier. The theory admits a de Sitter type solution, with healthy cosmological perturbations. We will show that there is 6 degrees of freedom propagate on top of de Sitter space-time, two tensor polarizations and four degrees of freedom related to two massless vector fields interacting with each other via Chern-Simons interaction term. We also investigate in detail the behavior of the geometric and physical parameters of a homogeneous and anisotropic Bianchi type I Universe, by using both analytical and numerical methods, by assuming that the matter content of the Universe can be described by the stiff causal and pressureless dust fluid equations of state.  The time evolution of the Bianchi type I Universe strongly depends on the initial conditions of the physical and geometrical quantities, as well as on the numerical values of the model parameters. Two important observational parameters, the mean anisotropy parameter, and the deceleration parameter, are also studied in detail, and we show that independently of the matter equation of state the cosmological evolution of the Bianchi type I Universe always ends in an isotropic and exponentially accelerating, de Sitter type, phase.

\end{abstract}

\maketitle

\section{Introduction}
One of the most interesting findings of modern cosmology is that our Universe is in a phase of accelerated expansion \cite{acc-papers}. To explain this fundamental observation,  one has to adopt a theory of gravity that admits a self-accelerating solution. Historically, such a solution was first obtained  by the addition of the cosmological constant to the Einstein gravitational field equations. But we now know that gravitational models assuming the existence of the   cosmological constant suffer from some phenomenological problems, such as, for example, the fine tuning problem \cite{cos-const-prob}. An alternative way to explain the recent acceleration of the Universe is to modify the dynamics of general relativity, such that the self-accelerated expansion is obtained from the  modified Einstein gravitational field equations. There are two ways to achieve this goal. The first one is to generalize the Einstein-Hilbert action, such as in $f(R)$ theories \cite{fRr}, or, more generally, in  $f(R,Q)$ theories where $Q$ can be any scalar combination of the curvature tensor/energy-momentum tensor \cite{fRRmunu}. The second possibility is enriching the graviton itself, such that it admits more than 2 degrees of freedom \cite{massive}, and constructing the massive gravity theory. In the latter case, the accelerated expansion of the Universe can be obtained not by imposing a repulsive gravity (like the cosmological constant), but with the reduction of attractive gravity.

An alternative way to produce a self-accelerated expanding Universe is to add some light degrees of freedom to the theory of general relativity. This can be considered as an addition of some matter degrees of freedom to the theory. The simplest possibility is to add a scalar field such that its equation of state mimics the equation of state of the cosmological constant, e.g. $p+\rho \approx 0$. This can be achieved by adding some higher derivative kinetic terms \cite{kessence}, or by adding a potential term of the scalar field to the theory \cite{quintessence}. Many works have been done in the context of scalar field cosmology, including inflationary \cite{inf} and dark energy models \cite{DE}. Recently, an interesting scalar field theory was proposed, which has higher than second order terms in the time derivative interactions in the action, although their equations of motion remains at most second order \cite{gali}. The cosmological implications of the Horndeski type theories have been  extensively investigated in the literature \cite{gali-cos}.

Another interesting possibility is to add a vector degree of freedom to the theory of Einstein general relativity. The simplest example is to add a Maxwell kinetic term to the Einstein-Hilbert action and obtain an Einstein-Maxwell system \cite{emax}. One can also consider a massive vector field as an Einstein-Proca system. Many works has been done in the context of Einstein-Maxwell and Einstein-Proca theories \cite{max-proca}. Another possibility is to consider the Weyl gravity, in which the metric compatibility condition does not hold anymore. As a result the covariant derivative of the metric tensor can be characterized by a vector field as $\nabla_\mu g_{\nu\rho}=w_\mu g_{\nu\rho}$ \cite{weyl-review,many-weyl}. One can also obtain the Einstein-Proca system from the higher dimensional Gauss-Bonnet theory in the context of Weyl geometry \cite{tomi}. Cosmological implications of such a theory has been considered in \cite{cos-weyl}. One can also think of a generalization of the Proca action (as in the Horndeski theory), in which the action has higher order derivatives, with at most second order time derivative terms in the equations of motion similar to the case of scalar galileon. But assuming the $U(1)$ symmetry, one can prove that such a vector galileon theory does not exist \cite{nogo}, meaning that the $U(1)$-vector galileon theory cannot exist. However, relaxing this constraint, one can generalize the Proca action, in a way that the helicity-0 of the vector field mimics the scalar galileon interactions \cite{vec-gali}. The cosmology of this vector galileon theory, as well as possible generalization, have been investigated  in the literature \cite{cos-vec-gali,gene-vec-gali}.

Other higher derivative interaction terms which are not included in the vector galileon theory may also be considered. These terms in general produce Ostrogradski instabilities that cause the propagation of ghost degrees of freedom at large scales. However, one can tune the coupling constants in a way that the ghost becomes non-dynamical at scales comparable to the Hubble radius. This will make the theory effectively reliable in these scales \cite{BP}.

 Other possibility of generalizing Einstein's general relativity within a purely geometric approach is to consider the effects of the torsion tensor \cite{cartan}. The torsion tensor can in general be decomposed into a vector field $Q_{\mu}$, a axial vector field $S^{\mu}$ and a tensor field $t_{\mu\nu\rho}$ satisfying the conditions
$t_{\mu\nu\rho}+t_{\nu\rho\mu}+t_{\rho\mu\nu}=0$, and $
g_{\mu\nu}t^{\mu\nu\rho}=0=g_{\mu\rho}t^{\mu\nu\rho}$, respectively.
Assuming that the tensor $t_{\mu\nu\rho}$ is zero, one can obtain a vector-tensor theory of gravity, which is well-known in the context of supergravity theory. The cosmological implications of the Gauss-Bonnet theory coupled to a Weyl vector in Cartan space-time were considered in \cite{WCGB}.

 An interesting vector-tensor theory of gravity is the Einstein-aether theory \cite{EA}. In this theory, one imposes a dynamical condition through a Lagrange multiplier in order to make the vector field always time-like. In this case the Universe has a predefined preferred time direction, which breaks the Lorentz invariance. Such a theory can explain the self-accelerated expansion of the Universe. The relation between the Einstein-aether theory and Horava-Lifshitz gravity \cite{horavarelation}, as well as its relation to the scalar-tensor theory \cite{SEA},  has been carefully investigated  in the physical literature.

 Scalar field dark energy models have provided a very successful description of the observational properties of the Universe, including the explanation of its late acceleration. However, \textit{a priori} one cannot simply reject
 the interesting idea that dark energy may have a more complex field
structure. One such possibility is to describe dark energy in terms of a vector or
Yang-Mills type field, which is also allowed to directly couple to gravity. The
simplest possible action for a Yang-Mills type dark energy model minimally coupled to gravity is \cite{v1}
\begin{eqnarray}
S_{V} &=&-\int d^{4}x\sqrt{-g}\Bigg\{\frac{R}{2}+\sum_{a=1}^{3}\Bigg[\frac{1%
}{16\pi }F_{\mu \nu }^{a}F^{a}{}^{\mu \nu }
+V(A^{2})\Bigg]+L_{m}\Bigg\},  \label{actv}
\end{eqnarray}%
where $F_{\mu \nu }^{a}=\nabla _{\mu }A_{\nu }^{a}-\nabla _{\nu }A_{\mu
}^{a} $, $a=1,2,3$,  $\nabla _{\mu }$ represents the covariant derivative with respect to the
metric, while $V(A^2)$ with $A^{2}=g^{\mu \nu }A_{\mu }^{a}A_{\nu }^{a}$ represents a self-interaction potential explicitly violating
gauge invariance. In the  action
given by Eq.~(\ref{actv}), the dark energy component is represented by three vector fields, and thus
the action (\ref{actv}) generalizes the Einstein-Maxwell single vector field dark energy model. The astrophysical and cosmological implications of the
vector type dark energy models and of their generalizations have been extensively investigated in \cite{v2}.

Extended vector field dark energy models either, in which
the vector field non-minimally couples to the gravitational field, have also been studied. Such
models have been proposed, and their cosmological properties have been investigated in, for example,   \cite{v3}. The action for
the non-minimally massive vector field coupled to gravity can be represented as
\begin{eqnarray}
S &=&-\int d^{4}x\sqrt{-g}\Bigg[\frac{R}{2}+\frac{1}{16\pi }F_{\mu \nu
}F^{\mu \nu }-\frac{1}{2}\mu _{\Lambda }^{2}A_{\mu }A^{\mu }
\label{1} +\omega A_{\mu }A^{\mu }R+\eta A^{\mu }A^{\nu }R_{\mu \nu }+L_{m}\Bigg],
\end{eqnarray}%
where $\mu _{\Lambda }$ is the mass of the massive cosmological vector
field, and $A^{\mu }\left( x^{\nu }\right) $, $\mu ,\nu =0,1,2,3$ is its
four-potential, which is allowed to couple non-minimally to gravity. Here $\omega $ and $\eta $
are dimensionless coupling parameters. By following a close analogy with
electrodynamics, the dark energy vector type field tensor is defined again as $F_{\mu \nu
}=\nabla _{\mu }A_{\nu }-\nabla _{\nu }A_{\mu }$.

A so called superconducting type
dark energy model was recently introduced in  \cite{SupracondDE}. Inspired by some condensed matter concepts, the starting
point of this approach is represented by the deep connection of the
gravitational actions for scalar field models, given by $S_{\phi }=-\int {d^{4}x\sqrt{-g}\left[ \frac{R}{2}-\frac{1}{2}\nabla
^{\alpha }\phi \nabla _{\alpha }\phi +V(\phi )\right] }$,
where $V(\phi )$ the self-interaction potential of the scalar field, and (\ref{actv}),
respectively. Despite having very different mathematical forms, the two actions for scalar and vector fields, respectively, can be interpreted and
described as the two limiting cases of a single unified fundamental physical
model that describes the spontaneous breaking of the U(1) symmetry of the
``electromagnetic" type dark energy, with the corresponding action given by
\begin{eqnarray}  \label{s1}
S&=&-\int \Bigg[ \frac{R}{2}+\frac{1}{16\pi }F_{\mu \nu }F^{\mu \nu }-\frac{%
\lambda}{2} g^{\mu \nu }
\left( A_{\mu }-\nabla _{\mu }\phi \right) \left( A_{\nu }-\nabla _{\nu
}\phi \right) +V\left( A^2,\phi \right) -
\frac{\alpha}{2} g^{\mu \nu }j_{\mu }\left( A_{\nu }-\nabla _{\nu }\phi
\right) +L_{m}\Bigg] \sqrt{-g}d^4x ,\nonumber\\
\end{eqnarray}%
where $\lambda $ and $\alpha $ are arbitrary constants, $L_m=L_{m}\left( g_{\mu \nu },\psi
\right) $ is the Lagrangian of the total (ordinary baryonic plus dark)
matter, and $j^{\mu }=\rho u^{\mu }$ is the total mass current. In Eq.~(\ref{s1}) $\rho $
denotes the total matter density (including the dark matter one), while $u^{\mu }$ is the
matter four-velocity.  Hence, as one can easily see,  the
gravitational action defined by Eq.~(\ref{s1}) provides a unified theoretical framework for the scalar-vector interactions in a gravitational background.

In the theory of electromagnetism, one can add to the electromagnetic Lagrangian a very interesting interaction term, known as the ``Chern-Simons" term, and  defined as \cite{undota}
\begin{align}  \label{01}
k_\mu\epsilon^{\mu\nu\rho\sigma}B_\nu B_{\rho\sigma},
\end{align}
where $B_{\mu\nu}=\partial_\mu B_\nu -\partial_\nu B_\mu$, and $k_\mu$ is a
constant vector acting as a coupling constant.
This term was initially defined as a topological mass term for gauge fields in (2+1) dimensions \cite{2p1}. One can see that because of the appearance of the Levi-Civita tensor, the CPT symmetry will be broken. On the other hand, because of a constant vector $k_\mu$, the Lorentz invariance will be broken by this term. This implies that the Chern-Simons term will reduce the Lorentz invariance symmetry to the 3 dimensional translational symmetry. Many works has been done in the context of Chern-Simons electrodynamics, including applications to quantum electrodynamics \cite{qed}. In \cite{epjc} the authors obtain the Chern-Simons term from dimensional reduction of the Carroll-Field-Jackiw Higgs model. Also in \cite{bound} the Chern-Simons theory with boundaries has been considered in more details.

 The Chern-Simons  modified gravity represents an interesting modification of general relativity, in which the Einstein-Hilbert action is extended by adding a parity-violating Chern-Simons term \cite{CS1,CS2,CS3}. This term couples to gravity via a scalar field. The Chern-Simons correction enhances parity violation through a pure curvature term, as opposed to the matter term, as is  considered in standard general relativity. It is important to note that Chern-Simons modified gravity can be obtained explicitly from superstring theory, where upon four-dimensional compactification the Chern-Simons term, appearing the Lagrangian density, plays an essential role due to the Green-Schwarz anomaly-canceling mechanism \cite{CS4}. Two distinct formulations of Chern-Simons modified gravity have been proposed, namely, the nondynamical formulation and the dynamical formulation (see \cite{CS3} for a review of the early results). In the first formulation, the Chern-Simons field is an arbitrary function, prescribed a priori, with its effective evolution equation being equivalent to a differential constraint on the space of its allowed solutions.  In the second approach, the Chern-Simons field is treated as a dynamical field, with its own effective stress-energy tensor, and obeying a dynamical evolution equation.  The possibility of observationally testing the dynamical Chern-Simons modified gravity by using the accretion disk properties around slowly-rotating black holes was considered in \cite{CS5}.  Specific signatures do appear in the electromagnetic spectrum, thus leading to the possibility of directly testing Chern-Simons modified gravity by using astrophysical observations of the emission spectra from accretion disks.

In this paper, we are going to investigate the cosmological implications of the Chern-Simons term as a representative of dark energy effect. To this end, we will promote the constant vector $k_\mu$ to a dynamical field and impose a constraint its norm remains constant by adding this property by a Lagrange multiplier. This will restore the Lorentz invariance of the theory. The Chern-Simons term will be generalized to
\begin{equation*}
\alpha\epsilon^{\mu\nu\rho\sigma}B_\mu A_\nu B_{\rho\sigma}+\lambda(A^\mu
A_\mu+\eta),
\end{equation*}
where $\lambda$ is a Lagrange multiplier,  $\eta$ and $\alpha$ are constant and $\epsilon^{\mu\nu\rho\sigma}$ is the Levi-Civita tensor. The new vector field $A_\mu$ plays the role of the constant vector $k_\mu$ in the theory and it is constrained to have a constant norm.
In the Minkowski background,
if one wants to quantize the resulting electrodynamics theory, one should assume that the
constant $k_\mu$ be purely spatial and has a form $k_\mu=(0,\vec{k})$ \cite{quant}. In this paper, we will not interested to the quantization of the theory and as a result we will not impose any constraint on the sign of the norm of the vector field $A_\mu$. Also, as we will be see in the following, in the case of FRW cosmology, one should assume that the vector field be time-like and has only $(t)$-component, because otherwise the spacial isotropy will be broken. However, for the sake of completeness, we will consider the case of anisotropic cosmological implications of the theory and assume that the vector field $A_\mu$ be space-like in this case.

The present paper is organized as follows. In the next section we will introduce the basic theoretical model, and obtain the necessary results for considering the cosmological implications of the model. In Section \ref{cos-1} we will consider the isotropic cosmology of the Chern-Simons model by obtaining the de Sitter solution, and performing the cosmological perturbations around it. In Section \ref{cos-2} we will assume that the vector $A_\mu$ becomes space-like, and then consider the case of anisotropic cosmology with the geometry described by the Bianchi type-I metric. The field equations and the basic physical parameters of the model are also introduced. In Section~\ref{sect5} we present a detailed numerical analysis of the evolution equations for different equations of state of the matter content of the Universe. In the last Section we discuss and conclude our results.

\section{The model}
We propose an action functional of the form
\begin{align}  \label{act}
S&=\int d^4x \sqrt{-g}\bigg[\kappa^2 R-\frac{1}{4 } A_{\mu\nu}A^{\mu\nu}-%
\frac{1}{4}B_{\mu\nu}B^{\mu\nu}+\alpha\epsilon^{\mu\nu\rho\sigma}B_\mu A_\nu
B_{\rho\sigma}+\lambda(A^\mu A_\mu+\eta) +V(A^2)\bigg]+S_m,
\end{align}
where $A_\mu$ and $B_\mu$ are two vectors fields, $A^2=A_\mu A^\mu$, $S_m$
is the matter action, $\eta$ is the norm of vector field $A_\mu$ with dimension of mass squared and $\lambda$ is the Lagrange multiplier which dynamically enforces that the vector field $A_\mu$ has a constant norm.
 This action consists of an Einstein-Maxwell system coupled to a constant norm vector field $A_\mu$. The two vector fields then interact via the topological Chern-Simons term.

Varying the action with respect to the Lagrange multiplier results in
\begin{align}  \label{1}
A_\mu A^\mu+\eta=0,
\end{align}
which states that the vector field $A_\mu$ should have a constant norm. The
equation of motion for $B_\mu$ can be written as
\begin{align}  \label{2}
\nabla_\beta B^{\beta\alpha}-\alpha\epsilon^{\alpha\beta\gamma\delta}B_\beta
A_{\gamma\delta}+2\alpha\epsilon^{\alpha\beta\gamma\delta}A_\beta
B_{\gamma\delta}=0.
\end{align}
The equation of motion for $A_\mu$ is
\begin{align}  \label{3}
\nabla_\beta A^{\beta\alpha}+2\lambda A^\alpha+2V^\prime
A^\alpha=\alpha\epsilon^{\alpha\beta\gamma\delta}B_\beta B_{\gamma\delta},
\end{align}
where we have defined $V^\prime=dV/dA^2$.
One can see that the $A_\mu$ field is sourced by the $B_\mu$ vector field
through the Chern-Simons term, and acquires an effective mass $2\lambda$.
Also $B_\mu$ vector field is sourced by the $A_\mu$ field through the
Chern-Simons term. In the case of vanishing $\alpha$, the two vector field
evolve independently.

The equation of motion for the metric is
\begin{align}  \label{4}
\kappa^2 G_{\alpha\beta}+&\frac{1}{8 } g_{\alpha\beta}A_{\mu\nu}A^{\mu\nu}-%
\frac{1}{2}A_{\mu\alpha}A^\mu_{~\beta}+\frac{1}{8 } g_{\alpha\beta}B_{\mu%
\nu}B^{\mu\nu}-\frac{1}{2}B_{\mu\alpha}B^\mu_{~\beta}+\lambda A_\alpha
A_\beta  \notag \\
&-\frac{1}{2}\lambda g_{\alpha\beta}(A_\mu A^\mu+\eta)-\frac{1}{2 } V
g_{\alpha\beta}+V^\prime A_\alpha A_\beta=\frac{1}{2}T_{\alpha\beta}.
\end{align}
In order to obtain the conservation of energy-momentum tensor in this model, first note that from the constraint equation \eqref{1}, one has $A^\mu\nabla_\alpha A_\mu=0$. Also, by taking the divergence of equation \eqref{3}, one can obtain
\begin{align}\label{4.1}
A^\alpha\nabla_\alpha\lambda+\lambda\nabla_\alpha A^\alpha+V^\prime\nabla_\alpha A^\alpha=\f14\epsilon^{\alpha\beta\gamma\delta}B_{\alpha\beta}B_{\gamma\delta}.
\end{align}
Now, taking the derivative of equation \eqref{4}, we arrive at
\begin{align}\label{5}
\nabla^\alpha T_{\alpha\beta}=\f14\alpha\epsilon^{\mu\rho\gamma\delta}\Big[2B_\rho B_{\gamma\delta}A_{\mu\beta}+B_{\mu\rho}B_{\gamma\delta}A_\beta+2B_\rho A_{\gamma\delta}B_{\mu\beta}-4A_\rho B_{\gamma\delta}B_{\mu\beta}\Big].
\end{align}
This shows that the energy-momentum tensor is not conserved due to non-linear interaction between two vector fields.

\section{Isotropic Cosmology of the Maxwell-Chern-Simons gravity}\label{cos-1}

In this paper, we will consider the cosmology of Maxwell-Chern-Simons
gravity theory. We will assume that the potential term in the action %
\eqref{act} is
\begin{align}
V(A^2)=-2\kappa^2\Lambda+\beta A_\mu A^\mu A_\nu A^\nu,
\end{align}
where $\Lambda$ is the cosmological constant.

\subsection{de Sitter solution}

Let us assume that our Universe can be described by a flat FRW metric of the form
\begin{align}
ds^2=-dt^2+a^2(t)\big(dx^2+dy^2+dz^2\big).
\end{align}
Moreover, because we are interested in the self-accelerated solutions to the
theory, we assume that the energy-momentum tensor vanishes. In this case,
the most general ansatz for the vector field which respects the spatial
isotropy is
\begin{align}
&A_\mu=(A_0,0,0,0), \\
&B_\mu=(B_0,0,0,0).
\end{align}
From the above equation, one can see that the vector field $A_\mu$ should be
time-like in this case, and we will assume that $\eta=m^2>0$. For the
space-like vector field $\eta<0$, there is no FRW solution unless $A_\mu=0$
which contradicts equation \eqref{1}.

By substituting the above ansatz to the field equations \eqref{1}-\eqref{4},
one can see that the constraint equation \eqref{1} gives $A_0=m$. Also, the $%
B_\mu$ equation of motion \eqref{2} satisfies automatically in this case.
The equations of motion for the vector field $A_\mu$ and for the metric can
then be written as
\begin{align}
2& m(\lambda-2\beta m^2)=0, \\
-3&\kappa^2 H^2+\kappa^2\Lambda+\frac{3}{2}\beta m^4-m^2\lambda=0, \\
-2&\kappa^2\dot H-3\kappa^2 H^2 +\kappa^2 \Lambda -\frac{1}{2}\beta m^4=0,
\end{align}
which can be solved as
\begin{align}
\lambda=2\beta m^2,\qquad H^2=\frac{\Lambda}{3}-\frac{\beta m^4}{6\kappa^2}.
\end{align}
We will also assume that $B_0={\rm constant}$ for the sake of simplicity. In order
to have a consistent de Sitter solution, one should has $\beta<2\kappa^2%
\Lambda/m^4$. Note that, in the absence of cosmological constant $\Lambda$,
the coupling constant $\beta$ should be negative in order to have a de
Sitter solution.

\subsection{Cosmological perturbations}

In this section we will perform the cosmological perturbation analysis on
top of the de Sitter solution obtained in the previous section. For the
metric perturbation, we have
\begin{align}
ds^2=-(1+2\phi)dt^2+2a(S_i+\partial_i B)dtdx^i+a^2\big((1+2\psi)\delta_{ij}+\partial_i\partial_j E+\partial_{(i}F_{j)}+h_{ij}%
\big)dx^i dx^j,
\end{align}
where $\phi$, $\psi$, $E$ and $B$ are the scalar perturbations, $F_i$ and $S_i$ are the vector
perturbations with the property $\partial_i F_i=0=\partial_i S_i$, and $h_{ij}$ is
associated with the tensor perturbation which is transverse and traceless $%
h_{ii}=0=\partial_i h_{ij}$. Also, the spatial indices are raised and lowered by $\delta_{ij}$.

For the vector fields, we decompose the
perturbed vector field as
\begin{align}
&A_\mu=(A_0+\delta A_0,\xi_i+\partial_i \delta A), \\
&B_\mu=(B_0+\delta B_0,\epsilon_i+\partial_i \delta B),
\end{align}
where $\delta A_0$, $\delta B_0$, $\delta A$ and $\delta B$ are the scalar perturbations
and $\xi_i$ and $\epsilon_i$ are the vector perturbations with the property $%
\partial_i\xi_i=0=\partial_i\epsilon_i$. The perturbation of the
Lagrange multiplier is $\lambda=\lambda_0+\delta\lambda$.

Under the general linear coordinate transformation $x^\mu\rightarrow x^\mu+\delta x^\mu$, the metric perturbations transforms as
\begin{subequations}
\begin{align}
\phi&\rightarrow\phi-\partial_t\delta x^0,\\
B&\rightarrow B+\f{1}{a}\delta x^0-a\partial_t\delta x,\\
\psi&\rightarrow\psi-H\delta x^0,\\
E&\rightarrow E-\delta x,\\
S_i&\rightarrow S_i-a\partial_t\eta_i,\\
F_i&\rightarrow F_i-2\eta_i,\\
h_{ij}&\rightarrow h_{ij},
\end{align}
\end{subequations}
where we have decomposed $\delta x^\mu$ to $(\delta x^0,\eta_i+\partial_i\delta x)$, with $\partial_i\eta_i=0$. Also the various components of a vector field $A_\mu$ will be transformed as
\begin{subequations}
\begin{align}
\delta A_0&\rightarrow \delta A_0-A_0\partial_t\delta x^0,\\
\delta A&\rightarrow \delta A-A_0\delta x^0,\\
\xi_i&\rightarrow\xi_i.
\end{align}
\end{subequations}
Likewise, we have a similar transformation rule for the vector field $B^\mu$. Also, one can find that the Lagrange multiplier will not change under this transformation
\begin{align}
\delta\lambda\rightarrow\delta\lambda.
\end{align}
We should note that in obtaining the above relations we have assumed that the background values $A_0$, $B_0$ and $\lambda_0$ are constant, as is obtained from the previous section.

With the aid of above transformations, one can construct 7 independent gauge invariant scalar perturbations, 3 independent gauge invariant vector perturbations and one gauge invariant tensor perturbation as follows
\begin{align}
\Phi&=\phi+\partial_t\left(aB-\f{a^2}{2}\partial_t E\right),\qquad\qquad
\Psi=\psi+H\left(aB-\f{a^2}{2}\partial_t E\right),\\
\delta \mathcal{A}_0&=\delta A_0+A_0\partial_t\left(aB-\f{a^2}{2}\partial_t E\right),\qquad
\delta \mathcal{A}=\delta A+A_0\left(aB-\f{a^2}{2}\partial_t E\right),\\
\delta \mathcal{B}_0&=\delta B_0+B_0\partial_t\left(aB-\f{a^2}{2}\partial_t E\right),\qquad
\delta \mathcal{B}=\delta B+B_0\left(aB-\f{a^2}{2}\partial_t E\right),
\end{align}
and the perturbation of the Lagrange multiplier is already gauge invariant. The vector perturbations $\xi_i$ and $\epsilon_i$ are already gauge invariant and the remaining gauge invariant vector perturbation can be constructed as
\begin{align}
\beta_i=S_i-\f12a\partial_t F_i.
\end{align}
Moreover the tensor perturbation $h_{ij}$ is already gauge invariant and we end up with 11 gauge invariant perturbation variables.

After substituting the above expressions to \eqref{act} and expand the
resulting action up to second order in perturbations, one can see that the
tensor, vector and scalar parts is decoupled completely from each other. In
the following we will consider them separately.

\subsubsection{Tensor perturbation}

The tensor perturbation $h_{ij}$ is transverse and traceless, and can be
described by two polarization modes $h_{+}$ and $h_\times$. After Fourier
decomposition, one can obtain the second order action of tensor perturbation
as
\begin{align}  \label{tenpert}
S^{(2)}_{tensor}=\frac{1}{2 } \kappa^2\int dtd^3\vec{k} a^3\sum_{\lambda=+,%
\times}\left(|\dot{h}_\lambda|^2-\frac{\vec{k}^2}{a^2}|h_\lambda|^2\right),
\end{align}
where $\vec{k}$ is the comoving wave vector and dot represents time derivative. Also, it is understood that in any term in the second order action, the argument of one of the perturbation variables is $k$, and for the other variable is $-k$, due to the definition of Dirac delta function. One can see that the
Chern-Simons coupling term does not contribute to the tensor perturbation.
The action \eqref{tenpert} is equivalent to that of Einstein-Hilbert theory.
So, the theory has two tensor propagating modes associated to the massless
graviton.

\subsubsection{Vector perturbation}

We have 3 transverse independent gauge invariant vector perturbations $\beta_i$, $\xi_i$ and $\epsilon_i$.
After expanding the action up to second order in perturbations and
performing the Fourier transformation, one can obtain
\begin{align}
S^{(2)}_{vector}=\frac{1}{2 } \int dtd^3\vec{k} a \sum_{i=1}^2\left[\kappa^2 \vec{k}^2 \beta_i^2 +\dot{\xi}_i^2-\frac{\vec{k}^2}{a^2}%
\xi_i^2 +\dot{\epsilon}_i^2-\frac{\vec{k}^2}{a^2}\epsilon_i^2+4\frac{i\alpha
B_0}{a^4}\vec{k}.\big(\vec{\xi}(-k)\times\vec{\epsilon}(k)\big)-4\frac{i\alpha
A_0}{a^4}\vec{k}.\big(\vec{\epsilon}(-k)\times\vec{\epsilon}(k)\big)\right],
\end{align}
where we have added the arguments of the last two terms for clarity.
The vector perturbation of the metric $\beta_i$ does not couple to the other
vector fields and is non-dynamical. Obtaining the equation of motion
of $\beta_i$ and substituting it back to the action, one can see that $\beta_i$ will
be vanishes from the action. We have left with 2 vector perturbations
associated with two vector fields $A_\mu$ and $B_\mu$. One can see that
these vector perturbations interact non-trivially with each other through
the Chern-Simons coupling, together with a self interaction of the vector field $B_\mu$. At the end, we will have 4 healthy vector degrees of freedom for the theory.

\subsubsection{Scalar perturbation}

In this section we will obtain the scalar perturbation part of the theory.
There are 5 scalar modes, corresponding to the metric, the vector fields and
the Lagrange multiplier. Expanding the action up to second order and
performing Fourier decomposition, one can write the action with gauge invariant quantities as
\begin{align}
S^{(2)}_{scalar}=&\int dtd^3\vec{k}a^3\Big[-6\kappa^2\dot{\Psi}^2+12\kappa^2
H\Phi\Psi+2\kappa^2\frac{\vec{k}^2}{a^2}\left(2\Phi\Psi+\Psi^2\right)+2%
\left(2\beta m^4-3\kappa^2 H^2\right)\Phi^2-8\beta m^3\Phi\delta \mc{A}_0  \notag
\\
&+\frac{1}{2}\frac{\vec{k}^2}{a^2}\left(\dot{\delta\mc{A}}^2-2\dot{\delta\mc{A}}\delta \mc{A}_0+\delta
\mc{A}_0^2\right)+4\beta m^2 \delta \mc{A}_0^2+\frac{1}{2}\frac{\vec{k}^2}{a^2}\left(%
\dot{\delta\mc{B}}^2-2\dot{\delta\mc{B}}\delta \mc{B}_0+\delta \mc{B}_0^2\right)-2m\delta
\mc{A}_0\delta\lambda+2m^2\Phi\delta\lambda\Big].
\end{align}
One can see from the above equation that $\delta\lambda$, $\delta \mc{B}_0$, $%
\delta \mc{A}_0$, and $\Phi$ are non-dynamical. Varying the action with respect
to these variables leads
\begin{align}
&2ma^3(m\Phi-\delta \mc{A}_0)=0, \\
&a\vec{k}^2(\delta \mc{B}_0-\dot \delta\mc{B})=0, \\
&2m\delta\lambda-\left(\frac{\vec{k}^2}{a^2}+8\beta m^2\right)\delta
\mc{A}_0+8\beta m^3\Phi+\frac{\vec{k}^2}{a^2}\dot{\delta\mc{A}}=0, \\
&m\delta\lambda+2\kappa^2\frac{\vec{k}^2}{a^2}\Psi+4\beta m^3(m\Phi-\delta
\mc{A}_0)+6\kappa^2 H(\dot{\Psi}-H\Psi)=0.
\end{align}
Solving the above equations for non-dynamical variables and substituting
them back to the action gives
\begin{align}  \label{act1}
S^{(2)}_{scalar}=\int dtd^3\vec{k}&\frac{a^3}{H\left(m^2\frac{\vec{k}^2}{a^2}%
-12\kappa^2 H^2\right)}\Big[2\kappa^2 H\frac{\vec{k}^2}{a^2}\left(m^2\frac{%
\vec{k}^2}{a^2}-\kappa^2\frac{\vec{k}^2}{a^2}-12\kappa^2H^2\right)\Psi^2
\notag \\
&+4\kappa^2\frac{\vec{k}^2}{a^2}\left(m^2\frac{\vec{k}^2}{a^2}-12\kappa^2
H^2\right)\Psi\dot{\Psi}+4\kappa^2m\frac{\vec{k}^2}{a^2}\Psi\Xi%
-6\kappa^2\frac{\vec{k}^2}{a^2}\Xi^2\Big],
\end{align}
where we have defined a new variable $\Xi=H\dot{\delta\mc{A}}-m\dot{\Psi}$. Varying the action with respect to $\Xi$ results in
\begin{align}
\Xi=\frac{m\kappa^2}{3Ha^2}\Psi.
\end{align}
Upon substituting the above relation to the action, one obtains
\begin{align}  \label{act1a}
S^{(2)}_{scalar}=\int dtd^3\vec{k}\frac{2\vec{k}^2\kappa^2}{3H^2a}\Psi%
\big[(\vec{k}^2+3H^2a^2)\Psi++6Ha^2\dot{\Psi}\big],
\end{align}
which can be solved for $\Psi$ with the result $\Psi=0$. So, the scalar part of the perturbed action vanishes and one can conclude that there is no dynamical scalar mode in the theory.

In summary, we have 2 gravitational wave mode and 4 vector modes associated with two vector fields in the theory. One should mention the Lagrange multiplier constraint implies that the vector field $A_\mu$ acquire only two degrees of freedom. This is similar to the case of Einstein-aether theory \cite{EA}.

\section{Anisotropic Cosmology: Bianchi type-I Universe}\label{cos-2}

In this section, we want to consider the case that the vector field $A_\mu$
is space-like. So, from now on we will set $\eta=-\mu^2$. In this case, at
least one of the spatial components of $A_\mu$ should be non-zero. For
simplicity, we choose the coordinate system such that the direction of the $%
x $ axis coincides with the direction of $A_\mu$. In this case the vector
field $A_\mu$ has the from
\begin{align}
A_\mu=(A_0(t),A_1(t),0,0),
\end{align}
In this case the vector field $B_\mu$ may have any direction in space and so
we set
\begin{align}
B_\mu=(B_0(t),B_1(t),B_2(t),B_3(t)).
\end{align}
The above ansatz for the vector field will break the spacial isotropy of the
space-time. In order to consider the cosmology of this case, let us assume
that the Universe is described by the Bianchi type-I metric of the form
\begin{align}  \label{6}
ds^2=-dt^2+\sum_{i=1}^{3}a_i^2(t)(dx^{i})^{2},
\end{align}
where $a_i$'s are the directional scale factors. We also assume that the
matter content of the Universe has a form
\begin{align}
T^{\mu}_\nu=\mathrm{diag}\big(-\rho(t),p_1(t),p_2(t),p_3(t)\big),
\end{align}
where $\rho$ is the energy density and $p_i$, $i=1,2,3$, is the pressure in
the direction $i$. Let us define the Hubble parameter, the deceleration
parameter and the anisotropy parameter as \cite{An}
\begin{align}  \label{33}
H_i&=\frac{\dot{a_i}}{a_i},i=1,2,3,\qquad V=a_1a_2a_3,\qquad H=\frac{1}{3}%
\sum_{i=1}^{3} H_i, \\
A&=\frac{1}{3}\sum_{i=1}^{3}\left(\frac{\Delta H_i}{H}\right)^2,\quad\quad%
\Delta H_i=H-H_i, i=1,2,3, \\
\Sigma^2&=\frac{1}{2}\left(\sum_{i=1}^{3} H_i^2-3H^2\right),\qquad q=\frac{d%
}{dt}\left(\frac{1}{H}\right)-1.
\end{align}
From Eqs.~(\ref{33}) we obtain immediately the important relation
\begin{equation}
\frac{\dot{V}}{V}=3H.
\end{equation}
The anisotropy parameter can be represented in an equivalent form as
\begin{equation}
A=\frac{1}{3H^2}\left(\sum _{i=1}^3{H_i^2}-3H^2\right),
\end{equation}
giving
\begin{equation}  \label{sum}
\sum _{i=1}^3{H_i^2}=3\left(1+A\right)H^2.
\end{equation}

Before solving the cosmological equations, we should note that the
Chern-Simons term can be considered as an effective cosmological constant,
as long as $A_{i}$'s and $B_{i}$'s do not change very rapidly. This is
because the Levi-Civita tensor is proportional to the volume of the Universe
$V$, so that the epsilon terms in the action varies very slowly, mimicking
the cosmological constant. However, from the vector field equations \eqref{2}
and \eqref{3}, one can see that the source terms (which are proportional to
the Chern-Simons term) decay at late times.

\subsection{Gravitational field equations}

Before writing the equations of motion of the theory, we should note that
the $(t)$-component of the vector field $B_{\mu }$ does not contribute to
the field equations. From now on, we will assume that $B_{0}=0$. On the
other hand, the $(t)$-component of the $A_{\mu }$ equation of motion %
\eqref{3}, implies that $A_{0}=0$. So, the constraint equation \eqref{1}
gives the remaining component of the vector field $A_{\mu }$ as $A_{1}=\mu
a_{1}(t).$ Also, the off-diagonal elements $(ij)$, $i\neq j$, of the metric
field equation gives $\dot{B}_{i}\dot{B}_{j}=0.$ This implies that at least
two out of three components of the vector field $B_{\mu }$ should be
constant. We will assume that $B_{1}$ and $B_{2}$ is constant. Now, the $(y)$%
-component of the $A_{\mu }$ field equation gives $B_{1}\dot{B}_{3}=0$,
which implies either $B_{1}=0$ or $B_{3}$ is constant. We will choose the
first possibility to make the vector field $B_{\mu }$ evolves in time. With
these in hand, the only remaining component of the $A_{\mu }$ field equation becomes
\begin{equation}
2\left( \lambda +2\beta \mu ^{2}\right) +H_{1}^{2}+2\frac{\alpha }{\mu }%
b_{2}(b_{3}H_{3}+\dot{b}_{3})-\frac{1}{V}\frac{d}{dt}\left( VH_{1}\right) =0.
\label{37}
\end{equation}%
The remaining components of the $B_{\mu }$ equation of motion are
\begin{equation}
2\left( b_{3}H_{3}+\dot{b}_{3}\right) +b_{3}H_{1}=0,  \label{38}
\end{equation}
or, equivalently,
\begin{equation}  \label{39a}
H_1=2H_3+2\frac{\dot{b_3}}{b_3},
\end{equation}
and
\begin{equation}
-2\mu \alpha b_{2}H_{1}+b_{3}H_{3}^{2}-3H\dot{b}_{3}-\ddot{b}_{3}-\frac{b_{3}%
}{V}\frac{d}{dt}\left( VH_{3}\right) =0.  \label{39}
\end{equation}
The Friedmann equation can be written as
\begin{equation}
\kappa ^{2}\left( 3\dot{H}+\sum_{i=1}^{3}H_{i}^{2}-\Lambda \right) +\frac{1}{%
4}\left( \frac{1}{4}b_{3}^{2}+\mu ^{2}\right) H_{1}^{2}-\frac{1}{2}\mu
^{2}(\lambda +\beta \mu ^{2})=-\frac{1}{4}(\rho +\sum_{i=1}^{3}p_{i}),
\label{40}
\end{equation}%
and the Raychaudhuri equations are
\begin{equation}  \label{42}
\kappa ^{2}\left( \frac{1}{V}\frac{d}{dt}\left( VH_{1}\right) -\Lambda
\right) -\frac{1}{4}\left( \frac{1}{4}b_{3}^{2}-\mu ^{2}\right) H_{1}^{2}+%
\frac{1}{2}\mu ^{2}(\lambda +3\beta \mu ^{2})=\frac{1}{4}(\rho
+p_{1}-p_{2}-p_{3}),
\end{equation}%
\begin{equation}  \label{43}
\kappa ^{2}\left( \frac{1}{V}\frac{d}{dt}\left( VH_{2}\right) -\Lambda
\right) -\frac{1}{4}\left( \frac{1}{4}b_{3}^{2}+\mu ^{2}\right) H_{1}^{2}-%
\frac{1}{2}\mu ^{2}(\lambda +\beta \mu ^{2})=\frac{1}{4}(\rho
-p_{1}+p_{2}-p_{3}),
\end{equation}%
and
\begin{equation}  \label{44}
\kappa ^{2}\left( \frac{1}{V}\frac{d}{dt}\left( VH_{3}\right) -\Lambda
\right) +\frac{1}{4}\left( \frac{1}{4}b_{3}^{2}-\mu ^{2}\right) H_{1}^{2}-%
\frac{1}{2}\mu ^{2}(\lambda +\beta \mu ^{2})=\frac{1}{4}(\rho
-p_{1}-p_{2}+p_{3}).
\end{equation}%
In the above equations we have defined
\begin{equation}
b_{2}(t)=\frac{B_{2}(t)}{a_{2}(t)},\qquad b_{3}(t)=\frac{B_{3}(t)}{a_{3}(t)}.
\end{equation}%
Equation \eqref{39a} can be solved for $b_{3}$ with the result
\begin{equation}
b_{3}(t)=\frac{c_{3}}{a_{3}\sqrt{a_{1}}},\qquad c_{3}=const.
\end{equation}%
Also from equation \eqref{5}, one can show that the matter field is conserved in this case
\begin{equation}
\dot{\rho}+3H\rho +\sum_{i=1}^{3}H_{i}p_{i}=0.
\end{equation}

\subsection{Deceleration parameter and anisotropy}

In the following we will restrict our analysis to the case of a
geometrically anisotropic Universe filled with a cosmological fluid with an
isotropic pressure distribution, with $p_{1}=p_{2}=p_{3}=p$.

By adding Eqs. (\ref{42})-(\ref{44}) we obtain
\begin{equation}
\kappa ^{2}\frac{1}{V}\frac{d}{dt}\left( 3HV\right) =3\kappa ^{2}\left( \dot{%
H}+3H^{2}\right) =\kappa^2 \frac{\ddot{V}}{V}=3\kappa ^{2}\Lambda +\frac{1}{4%
}\left( \frac{1}{4}b_{3}^{2}+\mu ^{2}\right) H_{1}^{2}+\frac{\mu ^{2}}{2}%
\left( \lambda -\beta \mu ^{2}\right) +\frac{3}{4}\left( \rho -p\right) .
\label{48}
\end{equation}

By substituting $3\dot{H}$ from the above equation into Eq. (\ref{40}) we
find the consistency condition
\begin{equation}
\kappa ^{2}\left( -9H^{2}+\sum_{i=1}^{3}H_{i}^{2}+2\Lambda \right) +\frac{1}{%
2}\left( \frac{1}{4}b_{3}^{2}+\mu ^{2}\right) H_{1}^{2}-\beta \mu ^{4}=-\rho
.  \label{consis}
\end{equation}

With the use of Eq.~(\ref{38}), Eq.~(\ref{37}) becomes
\begin{equation}
\frac{1}{V}\frac{d}{dt}\left( VH_{1}\right) =2\left( \lambda +2\beta \mu
^{2}\right) +H_{1}^{2}-\frac{\alpha }{\mu }b_{2}b_{3}H_{1}.  \label{50}
\end{equation}

By combining Eqs. (\ref{42}) and (\ref{50}) we find
\begin{equation}
\left[ 1-\frac{1}{4\kappa ^{2}}\left( \frac{1}{4}b_{3}^{2}-\mu ^{2}\right) %
\right] H_{1}^{2}-\frac{\alpha }{\mu }b_{2}b_{3}H_{1}-\Lambda +\lambda
\left( 2+\frac{\mu ^{2}}{2\kappa ^{2}}\right) +\beta \mu ^{2}\left( 4+\frac{%
3\mu ^{2}}{2\kappa ^{2}}\right) =\frac{1}{4\kappa ^{2}}(\rho -p).  \label{51}
\end{equation}

With the help of Eq.~(\ref{39a}), Eq.~(\ref{39}) can be reformulated as
\begin{equation}
\frac{1}{V}\frac{d}{dt}\left( VH_{3}\right) =\frac{1}{4}H_{1}^{2}-\frac{1}{%
b_{3}}\left( \dot{b}_{3}+2\mu \alpha b_{2}\right) H_{1}-\frac{1}{V}\frac{d}{%
dt}\left( V\frac{\dot{b}_{3}}{b_{3}}\right) ,  \label{52}
\end{equation}%
giving, after substitution into Eq.~(\ref{44}), the equation
\begin{equation}
\frac{1}{4}\left[ 1+\frac{1}{\kappa ^{2}}\left( \frac{1}{4}b_{3}^{2}-\mu
^{2}\right) \right] H_{1}^{2}-\frac{1}{b_{3}}\left( \dot{b}_{3}+2\mu \alpha
b_{2}\right) H_{1}-\frac{1}{V}\frac{d}{dt}\left( V\frac{\dot{b}_{3}}{b_{3}}%
\right) -\Lambda -\frac{1}{2\kappa ^{2}}\mu ^{2}(\lambda +\beta \mu ^{2})=%
\frac{1}{4\kappa ^{2}}(\rho -p)  \label{53}
\end{equation}

Adding Eqs. (\ref{51}) and (\ref{53}) we obtain
\begin{equation}
\frac{5}{4}H_{1}^{2}-\frac{1}{b_{3}}\allowbreak \left( \frac{\alpha }{\mu }%
b_{2}b_{3}^{2}+\dot{b}_{3}+2\mu \alpha b_{2}\right) H_{1}-\frac{1}{V}\frac{d%
}{dt}\left( V\frac{\dot{b}_{3}}{b_{3}}\right) -2\Lambda +\allowbreak \left(
4\beta \mu ^{2}+2\lambda +\beta \frac{\mu ^{4}}{\kappa ^{2}}\right) =\frac{1%
}{2\kappa ^{2}}(\rho -p),
\end{equation}%
while equating Eqs. (\ref{51}) and (\ref{53}) gives
\begin{equation}
\frac{1}{2}\left[ \frac{3}{2}-\f{1}{\kappa^2}\left( \frac{1}{4}b_{3}^{2}-\mu ^{2}\right) %
\right] H_{1}^{2}-\frac{1}{b_{3}}\left[ \frac{\alpha }{\mu }%
b_{2}b_{3}^{2}-\left( \dot{b}_{3}+2\mu \alpha b_{2}\right) \right] H_{1}+%
\frac{1}{V}\frac{d}{dt}\left( V\frac{\dot{b}_{3}}{b_{3}}\right) +\frac{1}{%
\kappa ^{2}}\left( 2\beta \mu ^{2}+\lambda \right) \left( 2\kappa ^{2}+\mu
^{2}\right) =0.
\end{equation}

From Eq. (\ref{48}) we obtain immediately the deceleration parameter as
\begin{equation}
q=2-\frac{1}{H^2}\left[ \Lambda +\frac{1}{12\kappa ^{2}}\left( \frac{1}{4}%
b_{3}^{2}+\mu ^{2}\right) H_{1}^{2}+\frac{\mu ^{2}}{6\kappa ^{2}}\left(
\lambda -\beta \mu ^{2}\right) +\frac{1}{4\kappa ^{2}}\left( \rho -p\right) %
\right] ,
\end{equation}%
while from Eq. (\ref{consis}) we obtain the anisotropy parameter in the form
\begin{equation}
A=2+\beta \frac{\mu ^{4}}{3H^2\kappa ^{2}}-\frac{2}{3}\frac{\Lambda }{H^{2}}-%
\frac{1}{6\kappa ^{2}}\frac{1}{H^{2}}\left( \frac{1}{4}b_{3}^{2}+\mu
^{2}\right) H_{1}^{2}-\frac{1}{3H^2\kappa ^{2}}\rho .
\end{equation}

\subsection{The evolution equations for the Bianchi type I cosmological model%
}

Since the field $b_{2}$ does not appear in the expressions of $q$ and $A$,
we can take $b_{2}=0$ without any lack of generality. Then from Eq. (\ref{51}%
) we obtain for $H_{1}$ the expression

\begin{equation}
H_{1}^{2}=\frac{\frac{1}{4\kappa ^{2}}(\rho -p)+\Lambda _{eff}}{1-\frac{1}{%
4\kappa ^{2}}\left( \frac{1}{4}b_{3}^{2}-\mu ^{2}\right) },
\end{equation}
where we have denoted
\begin{equation}
\Lambda _{eff}=\Lambda -\lambda \left( 2+\frac{\mu ^{2}}{2\kappa ^{2}}%
\right) -\beta \mu ^{2}\left( 4+\frac{3\mu ^{2}}{2\kappa ^{2}}\right) .
\end{equation}

By substituting this expression of $H_{1}^{2}$ into Eq. (\ref{48}) we obtain
the evolution equation for $V$ as
\begin{equation}
\frac{\ddot{V}}{V}=3\Lambda +\frac{\mu ^{2}}{2\kappa ^{2}}\left( \lambda
-\beta \mu ^{2}\right) +\frac{1}{4\kappa ^{2}}\frac{\left( \frac{1}{4}%
b_{3}^{2}+\mu ^{2}\right) }{1-\frac{1}{4\kappa ^{2}}\left( \frac{1}{4}%
b_{3}^{2}-\mu ^{2}\right) }\left[ \frac{1}{4\kappa ^{2}}(\rho -p)+\Lambda
_{eff}\right] +\frac{3}{4\kappa ^{2}}\left( \rho -p\right) .  \label{62}
\end{equation}

Eq. (\ref{53}) gives the time evolution of the field $b_{3}$ as
\begin{eqnarray}  \label{63}
\frac{1}{V}\frac{d}{dt}\left( V\frac{\dot{b}_{3}}{b_{3}}\right) &=&\frac{1}{4%
}\left[ 1+\frac{1}{\kappa ^{2}}\left( \frac{1}{4}b_{3}^{2}-\mu ^{2}\right) %
\right] \frac{\frac{1}{4\kappa ^{2}}(\rho -p)+\Lambda _{eff}}{1-\frac{1}{%
4\kappa ^{2}}\left( \frac{1}{4}b_{3}^{2}-\mu ^{2}\right) }-\frac{\dot{b}_{3}%
}{b_{3}}\sqrt{\frac{\frac{1}{4\kappa ^{2}}(\rho -p)+\Lambda _{eff}}{1-\frac{1%
}{4\kappa ^{2}}\left( \frac{1}{4}b_{3}^{2}-\mu ^{2}\right) }}  \notag \\
&&-\Lambda -\frac{1}{2\kappa ^{2}}\mu ^{2}(\lambda +\beta \mu ^{2})-\frac{1}{%
4\kappa ^{2}}(\rho -p).
\end{eqnarray}

\subsection{Dimensionless form of the cosmological evolution equations}

In order to simplify the mathematical formalism we rescale the physical and
geometrical quantities by introducing the set of dimensionless variables $%
\left( \tau ,r,P,B_{0},\mu _{0},\beta _{0},h\right) $, defined as
\begin{align}
&t=\frac{1}{\sqrt{3\Lambda +\frac{\mu ^{2}}{2\kappa ^{2}}\left( \lambda
-\beta \mu ^{2}\right) }}\tau,\quad H=\sqrt{3\Lambda +\frac{\mu ^{2}}{%
2\kappa ^{2}}\left( \lambda -\beta \mu ^{2}\right) }\,h,\nonumber\\& \rho =4\kappa
^{2}\Lambda _{eff}r,\quad p=4\kappa ^{2}\Lambda _{eff}P,\quad b_{3}=2\kappa B_{0},\quad \mu
=\kappa \mu _{0}, \quad\beta =\frac{\beta _{0}}{\kappa ^{2}}.
\end{align}

Then the system of Eqs. (\ref{62}) and (\ref{63}) becomes
\begin{equation}
\frac{1}{V}\frac{d^{2}V}{d\tau ^{2}}=1+\lambda _{1}\frac{\left(
B_{0}^{2}+\mu _{0}^{2}\right) }{1-\frac{1}{4}\left( B_{0}^{2}-\mu
_{0}^{2}\right) }\left[ (r-P)+1\right] +12\lambda _{1}\left( r-P\right) ,
\label{eqf1}
\end{equation}%
\begin{eqnarray}
\frac{1}{V}\frac{d}{d\tau }\left( V\frac{1}{B_{0}}\frac{dB_{0}}{d\tau }%
\right) &=&\lambda _{1}\left[ 1+\left( B_{0}^{2}-\mu _{0}^{2}\right) \right]
\frac{(r-P)+1}{1-\frac{1}{4}\left( B_{0}^{2}-\mu _{0}^{2}\right) }-2\sqrt{%
\lambda _{1}}\frac{1}{B_{0}}\frac{dB_{0}}{d\tau }\sqrt{\frac{(r-P)+1}{1-%
\frac{1}{4}\left( B_{0}^{2}-\mu _{0}^{2}\right) }}  \notag  \label{eqf2} \\
&&-4\lambda _{1}(r-P)-\lambda _{2},
\end{eqnarray}%
where
\begin{equation}\label{lambda1}
\lambda _{1}=\frac{\Lambda -\lambda \left( 2+\frac{\mu _{0}^{2}}{2}\right)
-\beta _{0}\mu _{0}^{2}\left( 4+\frac{3\mu _{0}^{2}}{2}\right) }{4\left[
3\Lambda +\frac{\mu _{0}^{2}}{2}\left( \lambda -\beta _{0}\mu
_{0}^{2}\right) \right] },
\end{equation}%
and
\begin{equation}\label{lambda2}
\lambda _{2}=\frac{\Lambda +\frac{1}{2}\mu _{0}^{2}(\lambda +\beta _{0}\mu
_{0}^{2})}{3\Lambda +\frac{\mu _{0}^{2}}{2}\left( \lambda -\beta _{0}\mu
_{0}^{2}\right) }.
\end{equation}%
The energy conservation equation can be written as
\begin{equation}
\frac{dr}{d\tau }+3h\left( r+P\right) =0.
\end{equation}%
By assuming an equation of state of the form
\begin{equation}
P=(\gamma -1)r,\qquad\gamma =\mathrm{constant},\qquad 1\leq \gamma \leq 2,
\end{equation}%
we obtain for the energy density of the cosmological matter the expression
\begin{equation}
r=\frac{r_{0}}{V^{\gamma }},
\end{equation}%
where $r_{0}$ is an arbitrary integration constant. The numerical value of $r_0$ can be taken as one without any loss of generality, so that the present density of the Universe, at the time $t=t_{pres}$,  is given by $\rho \left(t_{pres}\right)=4\kappa ^2\Lambda _{eff}/V^{\gamma }\left(t_{pres}\right)$. Therefore the system of
equations (\ref{eqf1}) and (\ref{eqf2}), describing the evolution of a
Bianchi type I Universe in the Maxwell-Chern-Simons theory can be
reformulated as a first order dynamical system given by
\begin{equation}
\frac{dV}{d\tau }=u,  \label{de1}
\end{equation}%
\begin{equation}
\frac{du}{d\tau }=\left\{ 1+\lambda _{1}\frac{\left( B_{0}^{2}+\mu
_{0}^{2}\right) }{1-\frac{1}{4}\left( B_{0}^{2}-\mu _{0}^{2}\right) }\left[
\frac{\left( 2-\gamma \right) r_{0}}{V^{\gamma }}+1\right] +12\lambda _{1}%
\frac{\left( 2-\gamma \right) r_{0}}{V^{\gamma }}\right\} V,  \label{de2}
\end{equation}%
\begin{equation}
\frac{dB_{0}}{d\tau }=b_{0},  \label{de3}
\end{equation}%
\begin{eqnarray}  \label{de4}
\frac{db_{0}}{d\tau } &=&\frac{1}{B_{0}}b_{0}^{2}+\lambda _{1}\left[
1+\left( B_{0}^{2}-\mu _{0}^{2}\right) \right] \frac{\left( 2-\gamma \right)
r_{0}V^{-\gamma }+1}{1-\frac{1}{4}\left( B_{0}^{2}-\mu _{0}^{2}\right) }%
B_{0}-\left[ 2\sqrt{\lambda _{1}}\sqrt{\frac{\left( 2-\gamma \right)
r_{0}V^{-\gamma }+1}{1-\frac{1}{4}\left( B_{0}^{2}-\mu _{0}^{2}\right) }}+%
\frac{u}{V}\right] b_{0}  \notag \\
&&-\left[ 4\lambda _{1}\frac{\left( 2-\gamma \right) r_{0}}{V^{\gamma }}%
+\lambda _{2}\right] B_{0}.
\end{eqnarray}

The system of differential equations (\ref{de1})-(\ref{de4}) must be
integrated with the initial conditions $V(0)=V_{0}$, $u(0)=u_{0}$, $%
B_{0}(0)=B_{0}^{(0)}$, and $b_{0}(0)=b_{0}^{(0)}$, respectively. The
behavior of the solutions of the dynamical system is determined by two
arbitrary parameters, $\lambda _{1}$ and $\lambda _{2}$, representing the
combination of the four free model parameters $\left( \Lambda ,\lambda
,\beta ,\mu \right) $.

The deceleration parameter can be written as
\begin{equation}
q=2-9\frac{V^{2}}{u^{2}}\left[ \f{1}{3}+\frac{1}{3}\lambda _{1}\left(
B_{0}^{2}+\mu _{0}^{2}\right) \frac{\left( 2-\gamma \right) r_{0}V^{-\gamma
}+1}{1-\frac{1}{4}\left( B_{0}^{2}-\mu _{0}^{2}\right) }+4\lambda _{1}\frac{%
\left( 2-\gamma \right) r_{0}}{V^{\gamma }}\right] .
\end{equation}

From Eq. (\ref{40}) we obtain the following relation between the
deceleration and the anisotropy parameters
\begin{equation}
A=q+\frac{3V^{2}}{u^{2}}\left\{\lambda_2-\lambda _{1}\frac{\left( B_{0}^{2}+\mu
_{0}^{2}\right) }{1-\frac{1}{4}\left( B_{0}^{2}-\mu _{0}^{2}\right) }\left[
\left( 2-\gamma \right) r_{0}V^{-\gamma }+1\right] -4\left( 3\gamma -2\right)
\lambda _{1}\frac{r_{0}}{V^{\gamma }}\right\},
\end{equation}
Hence the behavior of both $q$ and $A$ is determined by the set of model
parameters $\left( \mu _{0},\lambda _{1},\lambda_2,r_{0}\right) $.

\section{Cosmological evolution of the Bianchi type I Universes in the
	Maxwell-Chern-Simons theory}\label{sect5}

In the present Section we will investigate the time dependence of the
geometrical and thermodynamical parameters of the Bianchi type I space-times
in the Maxwell-Chern-Simons theory. In order to obtain a relevant physical
and cosmological picture we adopt for the description of the matter content
a number of equations of state that could be relevant for the understanding
of the properties of the ultra-high density matter the Universe may have
contained in its very early stages.

In order to numerically integrate the evolution equations (\ref{de1})-(\ref%
{de4}) we will fix the numerical values of the parameters $\left(\lambda _1,\lambda _2\right)$. Once this is done, the numerical values of
the free parameters of the model, $\mu _0$ and $\beta _0$, can
be obtained from Eqs.~(\ref{lambda1}) and (\ref{lambda2}) as
\be
\mu _0=\frac{\sqrt{2-6 \lambda _2}}{\sqrt{4 \lambda _1+2 \lambda _2-1}},
\ee
and
\be
\beta _0=\frac{\lambda \left(4 \lambda _1+2\lambda _2-1\right)  \left[-\lambda _2+\lambda +\Lambda  \left(4 \lambda _1+2 \lambda _2-1\right)\right]}{6 \lambda _2^2+4
   \lambda _2-2},
\ee
respectively. It is interesting to note that the numerical value of the coefficient $\mu _0$ is determined by $\lambda _1$ and $\lambda _2$ only, while $\beta _0$ is also determined by the arbitrary values of $\lambda $.

\subsection{Stiff Fluid Filled Bianchi Type I Universe}

An important equation of state, extensively used to describe the properties
of high density matter, is the Zeldovich (or stiff matter) equation of
state, which can be used for matter densities significantly
higher than nuclear densities, $\rho >10\rho _{n}$, where $\rho _{n}$ is the
nuclear density. The Zeldovich equation of state can be obtained
theoretically from a relativistic Lagrangian that allows bare nucleons to
interact attractively by exchanging a scalar meson, and to interact
repulsively by exchanging a massive vector meson \cite{Zel}. In the
non-relativistic limit both the quantum and classical description of strong
interactions yield Yukawa-type potentials. At the highest matter densities
the nuclear interactions are dominated by the vector meson exchange, and one
can show, by using a mean field approximation, that in the
extreme limit of infinite densities the pressure tends to the energy
density, $p\rightarrow \rho $. In this high density limit the speed of
sound, given by $c_{s}^{2}=dp/d\rho \rightarrow 1$. Therefore the stiff
fluid equation of state satisfies the causality
condition, which requires that the speed of sound is less than the speed of
light, $c_{s}\leq c$. For the Zeldovich fluid with $r=P$, the energy
conservation gives the dependence of the density as a function of the
comoving $V$ in the form $r=r_{0}/V^{2}$.

In the case of the stiff fluid the cosmological evolution equations take the
simple form
\begin{equation}\label{78}
\frac{dV}{d\tau }=u,\qquad\frac{du}{d\tau }=\left\{ 1+\lambda _{1}\frac{\left(
	B_{0}^{2}+\mu _{0}^{2}\right) }{1-\frac{1}{4}\left( B_{0}^{2}-\mu
	_{0}^{2}\right) }\right\} V,
\end{equation}
\begin{equation}\label{79}
\frac{dB_{0}}{d\tau }=b_{0},\qquad\frac{db_{0}}{d\tau }=\frac{1}{B_{0}}b_{0}^{2}+%
\left[ \lambda _{1}\frac{1+\left( B_{0}^{2}-\mu _{0}^{2}\right) }{1-\frac{1}{%
		4}\left( B_{0}^{2}-\mu _{0}^{2}\right) }-\lambda _{2}\right] B_{0}-\left[2
\sqrt{\frac{\lambda _{1}}{1-\frac{1}{4}\left( B_{0}^{2}-\mu _{0}^{2}\right) }%
}+\frac{u}{V}\right] b_{0}.
\end{equation}

\begin{figure*}[h]
	\centering
	\includegraphics[width=8.5cm]{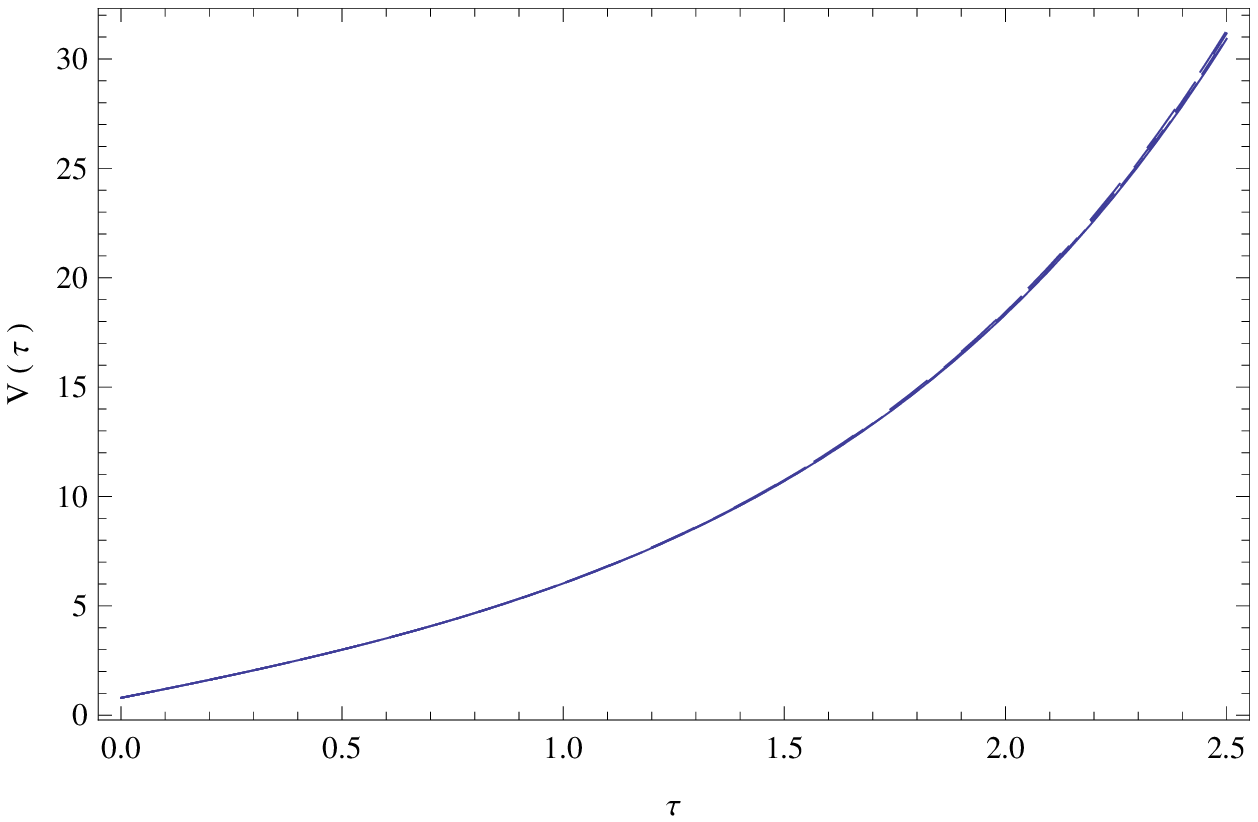}
	\includegraphics[width=8.5cm]{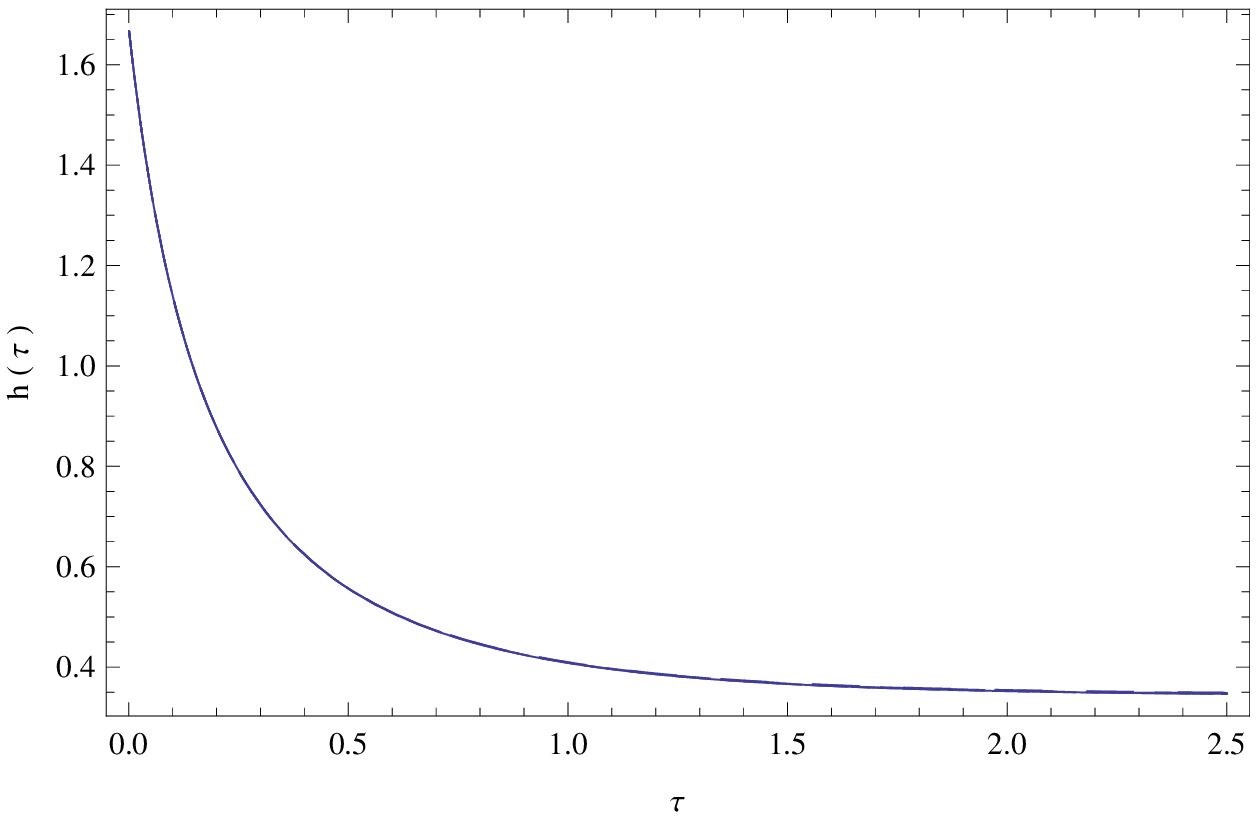}\\
	\includegraphics[width=8.5cm]{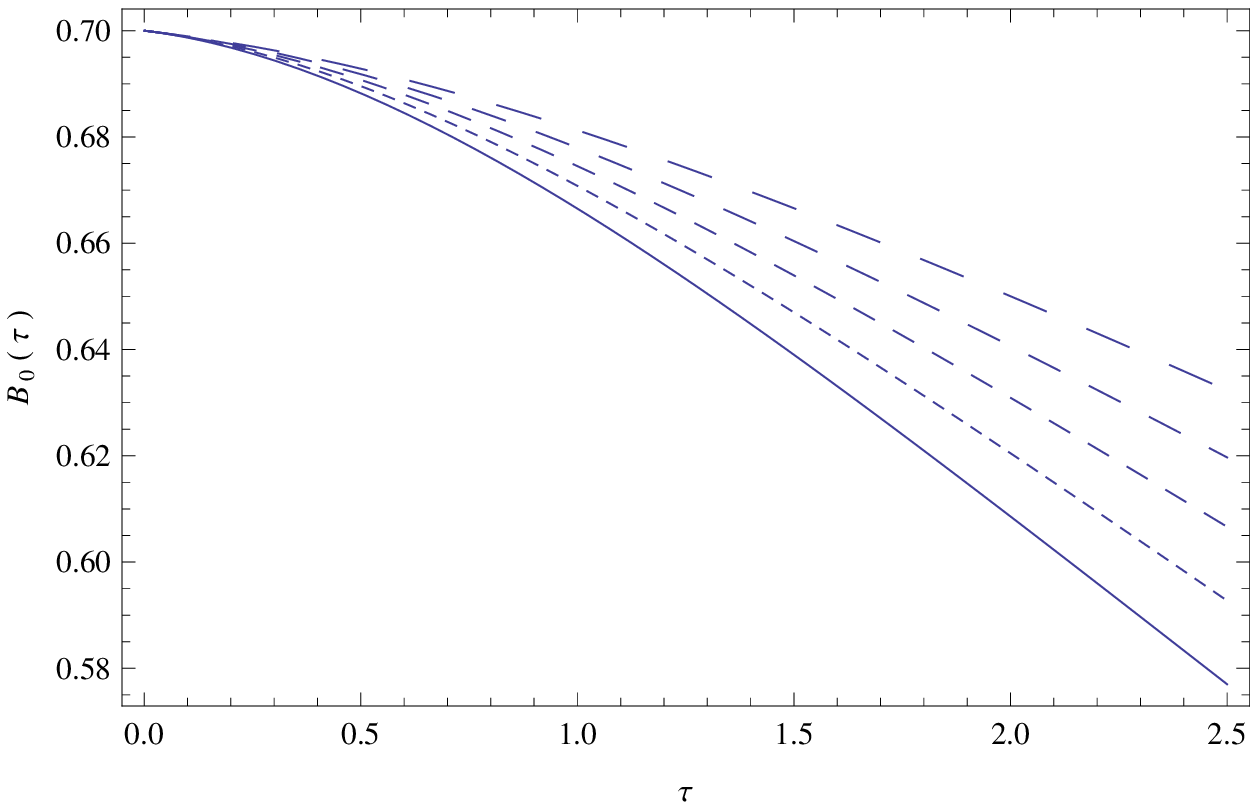}
	\includegraphics[width=8.5cm]{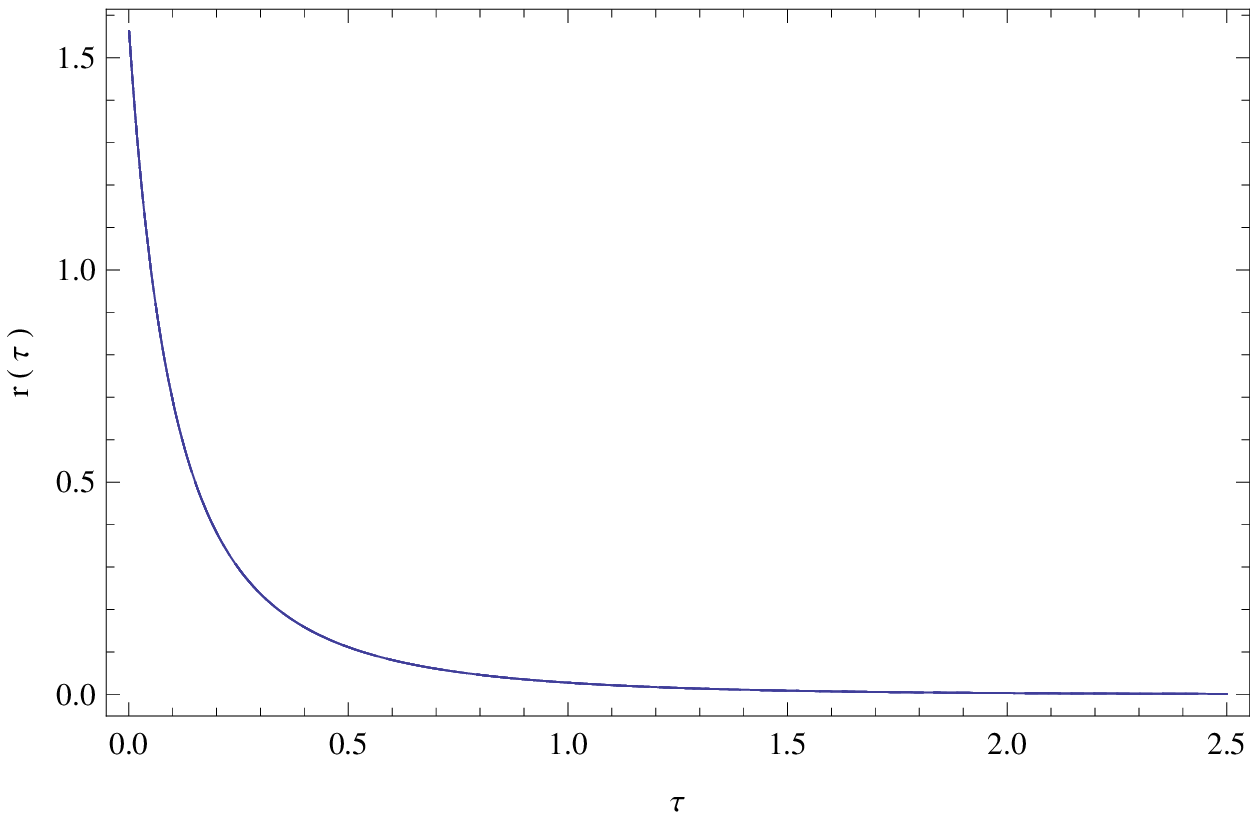}
	\caption{Variation of the comoving volume element $V$  (upper left figure), of
		the mean dimensionless Hubble function $h$ (upper right figure), of the dimensionless vector field $B_0$ (lower left figure), and of the dimensionless matter energy density $r$ for a Bianchi type I Universe filled with a stiff fluid in the Maxwell-Chern-Simons theory,  for different values of the parameters $\lambda _1$ and $\mu _0$: $\lambda _1=0.11$, $\mu _0= 0.447$ (solid curve), $\lambda _1 =0.12$, $\mu _0=0.377$ (dotted curve), $\lambda _1=0.13$, $\mu _0=0.333$ (short dashed curve), $\lambda _1 =0.14$, $\mu _0=0.301$ (dashed curve), and $\lambda _1 =0.15$, $\mu _0=0.277$ (long dashed
		curve), respectively.   The initial conditions used to numerically integrate the cosmological evolution equations are  $V (0)=0.8$, $u(0)=4$, $B_0 (0)=0.7$, and $b_0(0)=-0.001$, respectively. In all cases the numerical value of the parameter $\lambda _2$ has been fixed as  $\lambda _2=0.33$.}
	\label{fig1}
\end{figure*}

\begin{figure*}[h]
	\centering
	\includegraphics[width=8.5cm]{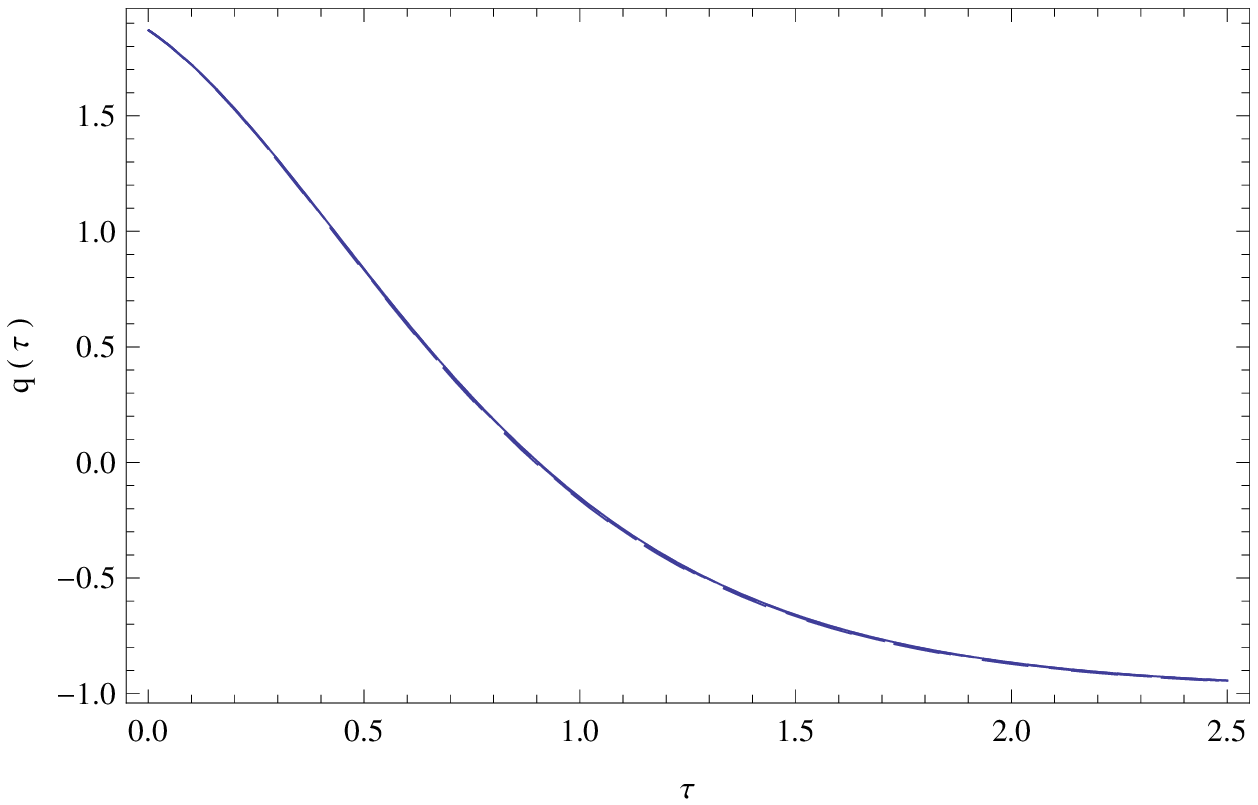}
	\includegraphics[width=8.5cm]{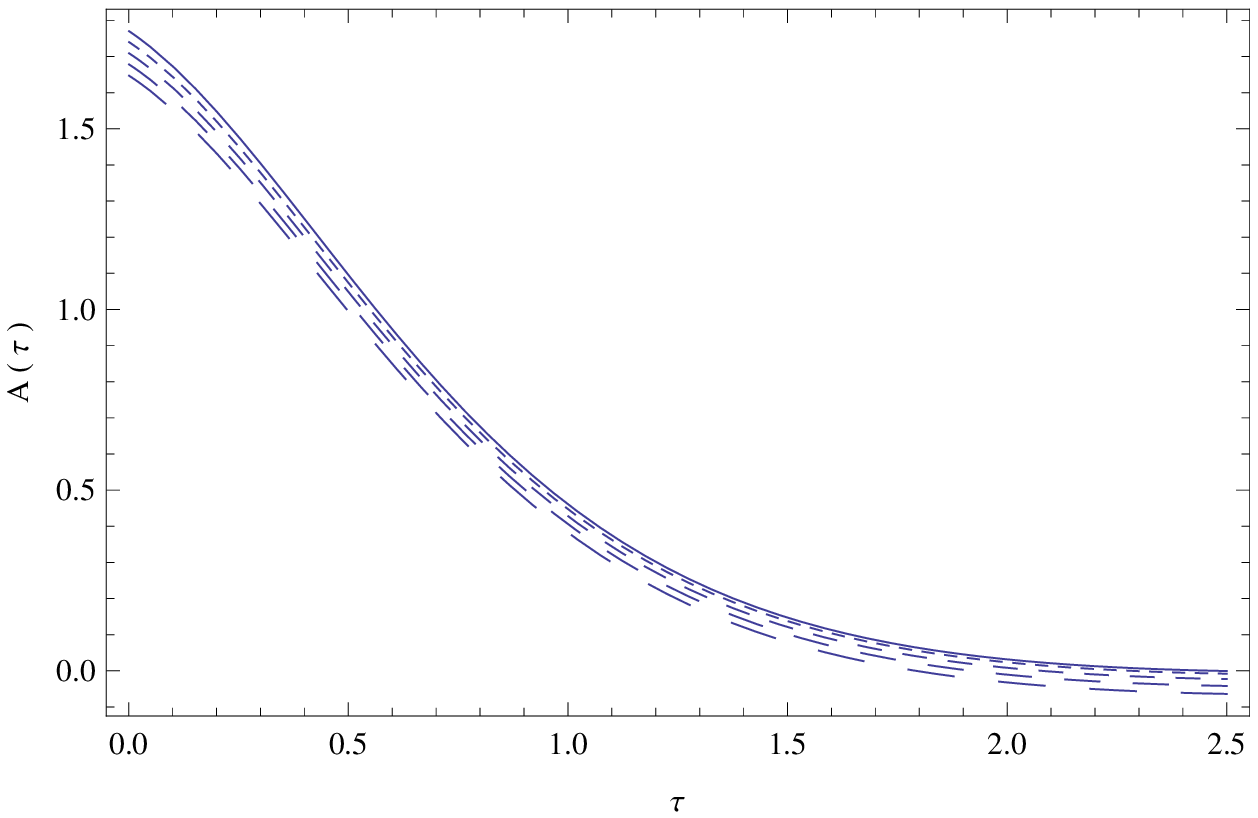}
	\caption{Variation of the deceleration parameter $q$  (left figure), and of
		the anisotropy parameter $A$ (right figure) for a Bianchi type I Universe filled with a stiff fluid in the Maxwell-Chern-Simons theory,   for different values of the parameters $\lambda _1$ and $\mu _0$: $\lambda _1=0.11$, $\mu _0= 0.447$ (solid curve), $\lambda _1 =0.12$, $\mu _0=0.377$ (dotted curve), $\lambda _1=0.13$, $\mu _0=0.333$ (short dashed curve), $\lambda _1 =0.14$, $\mu _0=0.301$ (dashed curve), and $\lambda _1 =0.15$, $\mu _0=0.277$ (long dashed
		curve), respectively.   The initial conditions used to numerically integrate the cosmological evolution equations are  $V (0)=0.8$, $u(0)=4$, $B_0 (0)=0.7$, and $b_0(0)=-0.001$, respectively. In all cases the numerical value of the parameter $\lambda _2$ has been fixed as  $\lambda _2=0.33$.}
	\label{fig2}
\end{figure*}

The results of the numerical integration of the above system are presented in Figs.~\ref{fig1} and \ref{fig2}, respectively.

In order to numerically integrate the cosmological evolution equations for the stiff fluid case we have fixed the value of the free parameter  $\lambda _2$ as $\lambda _2=0.33$. As initial conditions we have adopted the values $V(0)=0.8$, $u(0)=4$,  $B_0 (0)=0.7$, and $b_0(0)=-0.001$. The comoving volume element of the Bianchi type I Universe, presented in the left panel of Fig.~\ref{fig1}, is a monotonically increasing function of the cosmological time $\tau $, indicating an expansionary evolution of the Bianchi type I Universe. For small times the increase of $V$ is almost linear, and the variation of the numerical values of the parameter $\lambda _1$ influences the behavior of $V$ only in the large time limit. The mean Hubble function, shown in the right panel of Fig.~\ref{fig1}, is a monotonically decreasing function of time, and its behavior is slightly influenced in the large time limit by the variation of $\lambda _1$. In the large time limit $h$ tends to a constant value, indicating that in the presence of the Maxwell-Chern-Simons terms the stiff fluid filled Universe ends in an exponentially expanding de Sitter type era. The time evolution of the vector field $B_0$, depicted in the left lower panel of Fig.~\ref{fig1} shows a strong dependence on the numerical values of $\lambda _1$. $B_0$ is a monotonically decreasing function of time, and in the large time limit it takes very small numerical values. The energy density of the matter, represented in right lower panel of Fig.~\ref{fig1}, is a monotonically decreasing function of time, whose evolution is practically independent on the variation of the numerical values of $\lambda _1$. In the large time limit the matter energy density tends to zero, indicating that the de Sitter time expansion leads to a vacuum Universe, whose dynamics is dominated by the contributions of the Maxwell-Chern-Simons vector fields.

The time variations of the deceleration parameter of the Bianchi type I Universe, and of its anisotropy parameter, are represented in Fig.~\ref{fig2}.

The expansion of the Bianchi type I Universe begins in its very early stages with $q$ having values around $q\approx 2$. The Universe is initially in a decelerating state, with $q>0$, but for $\tau \approx 0.7$ the deceleration parameter reaches the value $q\approx 0$, and the Universe enters in an accelerating phase. In the large time limit $q$ reaches values of the order of $q\approx -1$, indicating the presence of the de Sitter expansion. The overall evolution of $q$ is slightly dependent on the  numerical values of $\lambda _1$. The time variation of the anisotropy parameter $A$, shown in the right panel of Fig.~\ref{fig2}, is also strongly dependent on the model parameter $\lambda _1$. In the large time limit $A\rightarrow 0$, indicating that when the Universe enters the de Sitter phase it is already in an isotropic state. High values of $\lambda _1$ lead to a rapid transition to the de Sitter era, as well as to the rapid isotropization of the Bianchi type I geometry.

\subsection{The dust Bianchi type I Universe}

As a next cosmological application of our model we will consider the case of the dust Universe, with the matter having negligibly small thermodynamic pressure, corresponding to the choice $\gamma =1$. The overall dynamics is very similar to the stiff fluid case, with the Bianchi I type Universe isotropizing in the large time limit, and ending in a de Sitter type expansionary phase. In the following we will investigate the effect of the variation of the parameter $\mu _0$ on the cosmological dynamics.

\begin{figure*}[h]
	\centering
	\includegraphics[width=8.5cm]{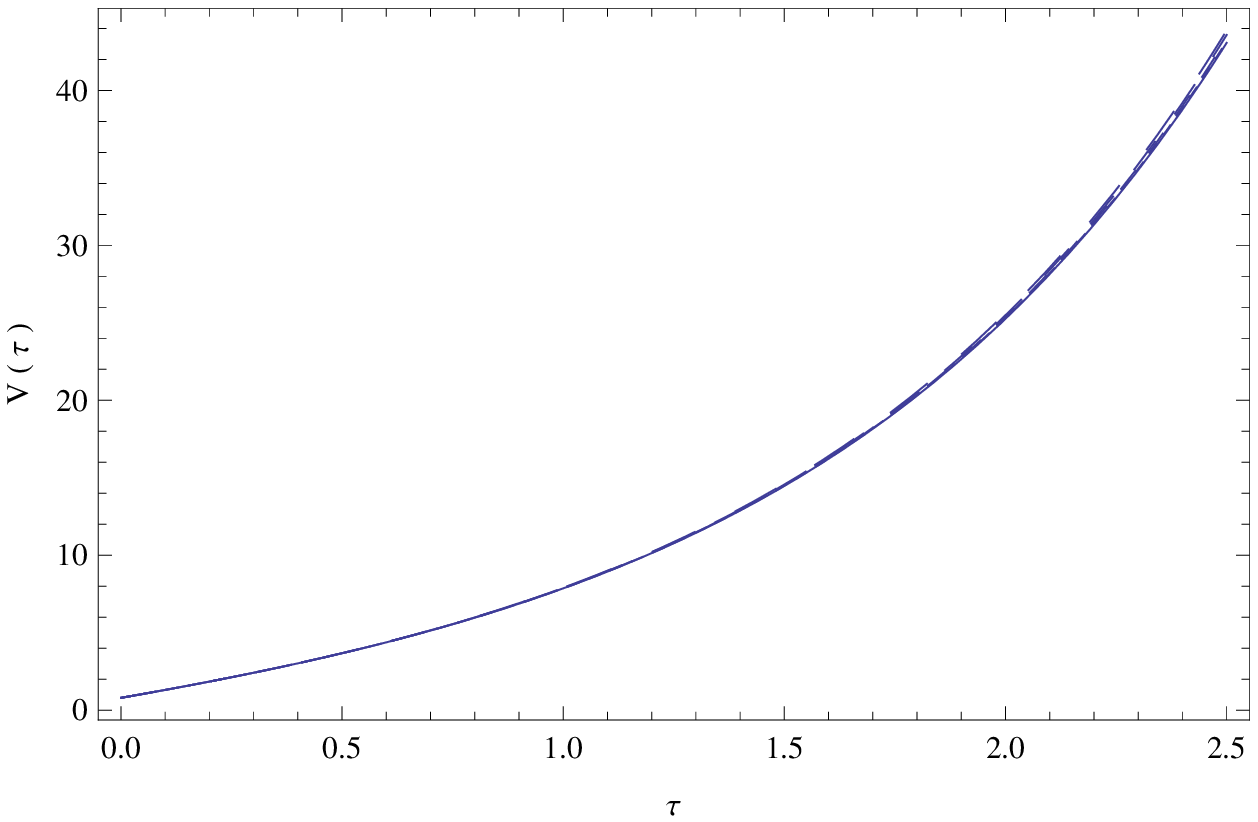}
	\includegraphics[width=8.5cm]{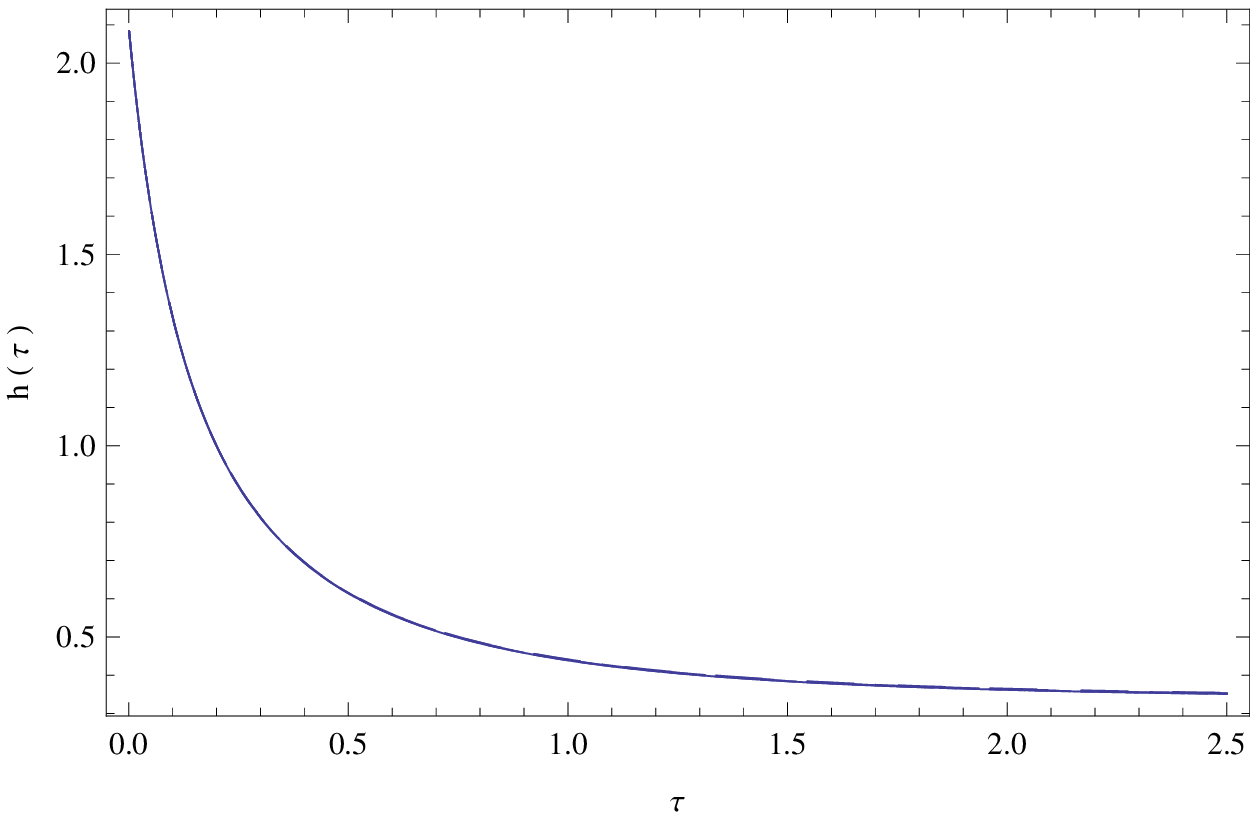}\\
	\includegraphics[width=8.5cm]{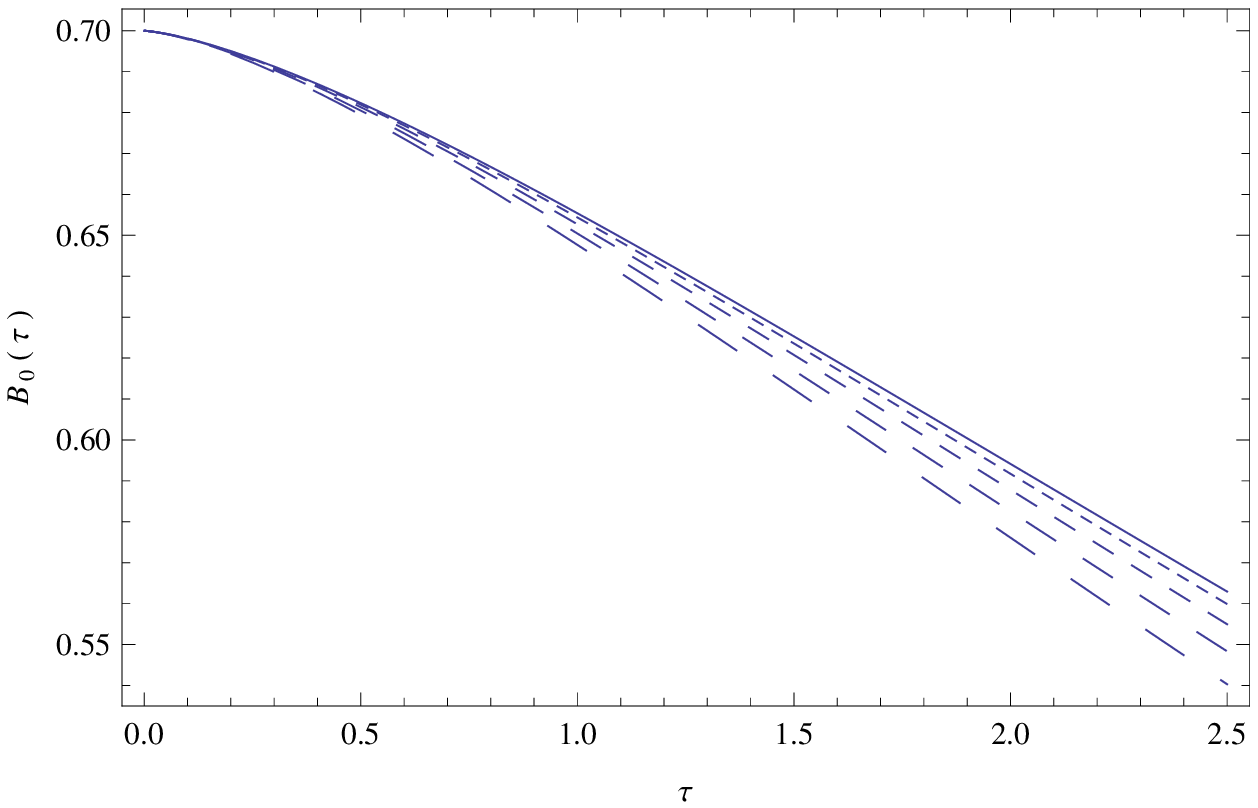}
	\includegraphics[width=8.5cm]{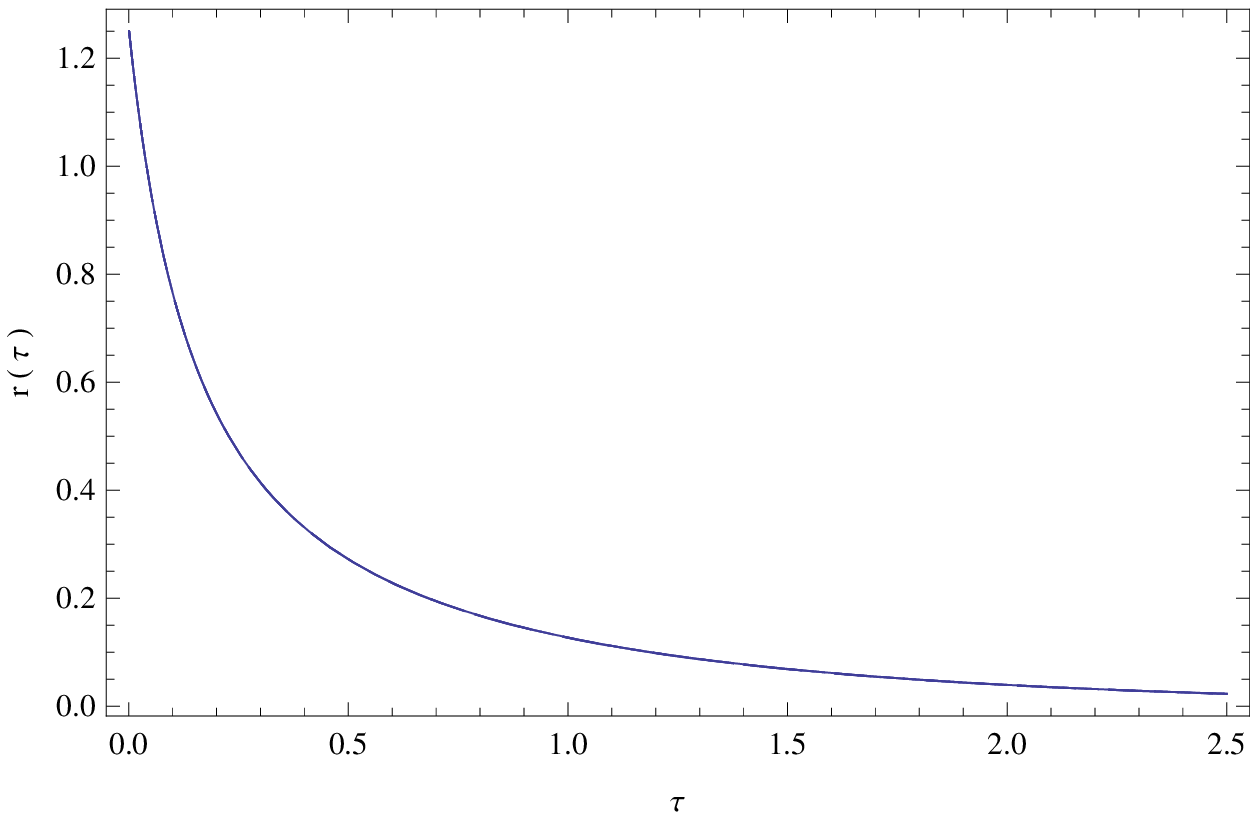}
	\caption{Variation of the comoving volume element $V$  (upper left figure), of
		the mean dimensionless Hubble function $h$ (upper right figure), of the dimensionless vector field $B_0$ (lower left figure), and of the dimensionless matter energy density $r$ for a dust fluid filled Bianchi type I Universe in the Maxwell-Chern-Simons theory,  for different values of the parameter $\mu _0$: $\mu _0 =0.10$ (solid curve), $\mu _0 =0.20$, (dotted curve), $\mu _0=0.30$, (short dashed curve), $\mu _0 =0.40$ (dashed curve), and $\mu _0 =0.50$ (long dashed
		curve), respectively.   The initial conditions used to numerically integrate the cosmological evolution equations are  $V (0)=0.8$, $u(0)=4$, $B_0 (0)=0.7$, and $b_0(0)=-0.001$, respectively. In all cases the value of the parameter $\lambda _1$ has been fixed as $\lambda _1=0.1$,  while $\lambda _2=\left(-4 \lambda _1 \mu _0^2+\mu _0^2+2\right)/2 \left(\mu _0^2+3\right)\approx 0.33$ for all considered values of $\mu _0$.}
	\label{fig3}
\end{figure*}

\begin{figure*}[h]
	\centering
	\includegraphics[width=8.5cm]{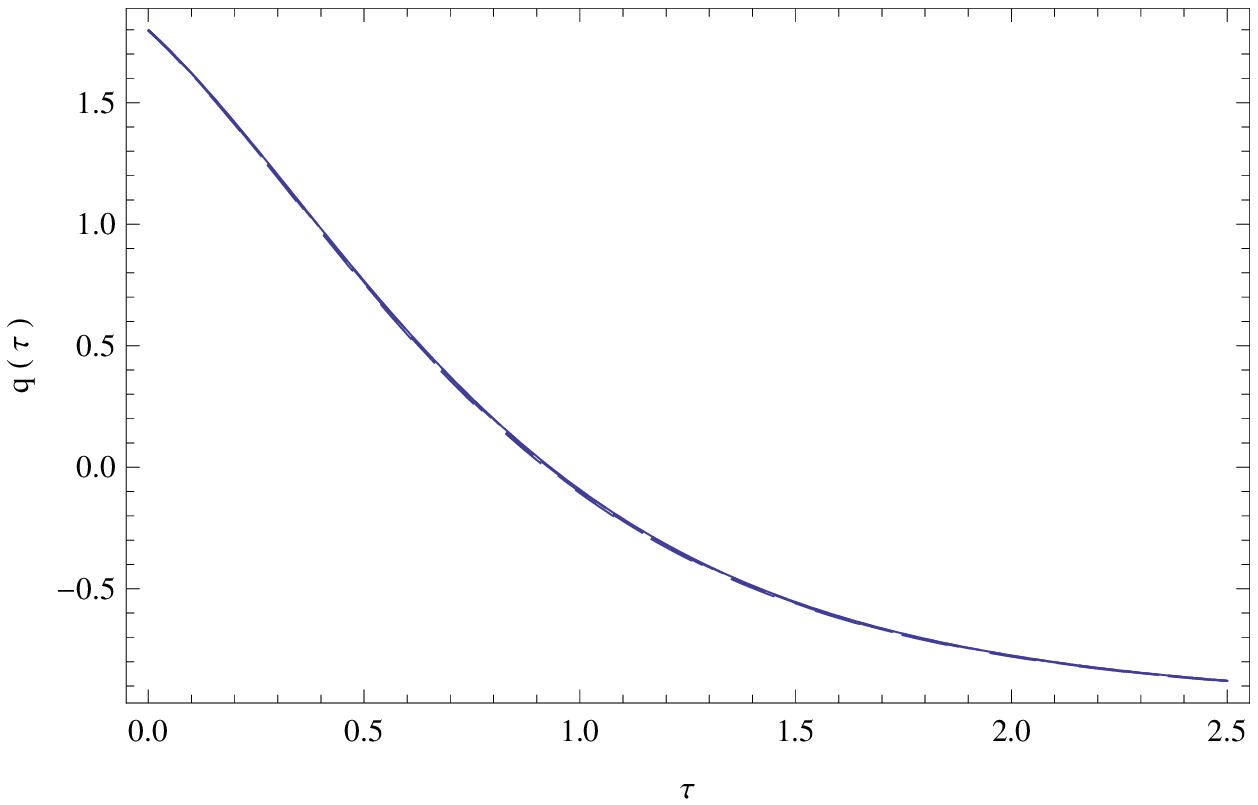}
	\includegraphics[width=8.5cm]{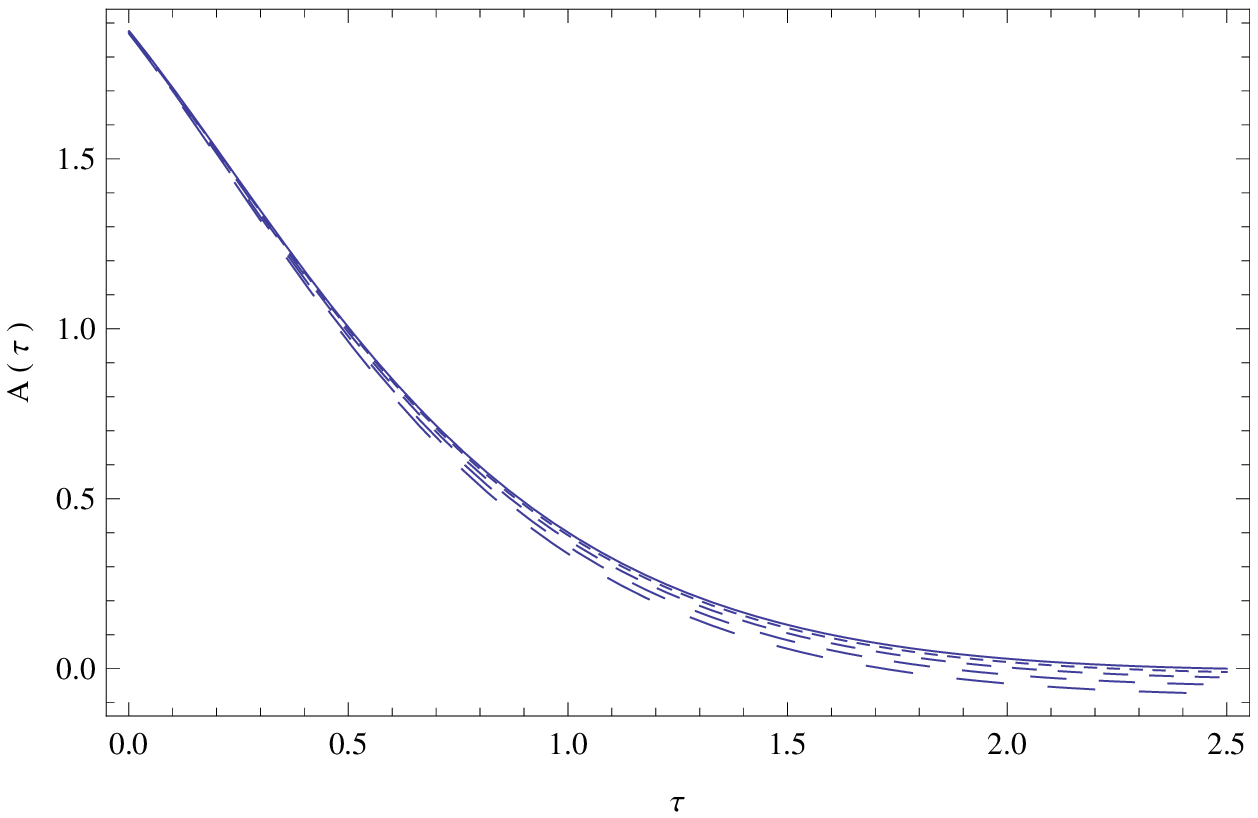}
	\caption{Variation of the deceleration parameter $q$  (left figure), and of
		the anisotropy parameter $A$ (right figure) for a dust fluid filled Bianchi type I Universe in the Maxwell-Chern-Simons theory,  for different values of the parameter $\mu _0$: $\mu _0 =0.10$ (solid curve), $\mu _0 =0.20$, (dotted curve), $\mu _0=0.30$, (short dashed curve), $\mu _0 =0.40$ (dashed curve), and $\mu _0 =0.50$ (long dashed
		curve), respectively.    The initial conditions used to numerically integrate the cosmological evolution equations are  $V (0)=0.8$, $u(0)=4$, $B_0 (0)=0.7$, and $b_0(0)=-0.001$, respectively. In all cases the value of the parameter $\lambda _1$ has been fixed as $\lambda _1=0.1$,  while $\lambda _2=\left(-4 \lambda _1 \mu _0^2+\mu _0^2+2\right)/2 \left(\mu _0^2+3\right)\approx 0.33$ for all considered values of $\mu _0$.}
	\label{fig4}
\end{figure*}

In order to numerically integrate the cosmological evolution equations in the case of the dust Universe we fix the numerical value of the free parameter $\lambda _1$ as $\lambda _1=0.1$,  and we vary the numerical values of $\mu _0$, and of $\lambda _2$, as given by
\be
\lambda _2=\frac{-4 \lambda _1 \mu _0^2+\mu _0^2+2}{2 \left(\mu _0^2+3\right)}.
\ee
Similarly to the stiff fluid case, as initial conditions we adopt the values $V(0)=0.8$, $u(0)=4$,  $B_0 (0)=0.7$, and $b_0(0)=-0.001$, respectively. The time variations of the comoving volume element, Hubble function, vector field, matter energy density, deceleration parameter and anisotropy parameter are plotted in Figs.~\ref{fig3} and \ref{fig4}, respectively.

The comoving volume element of the dust fluid filled Bianchi type I Universe, plotted in the left panel of Fig.~\ref{fig3}, is a monotonically increasing function of the cosmological time $\tau $, indicating that the Bianchi type I Universe is globally expanding. For small time intervals the variation of $V$ is almost linear. The mean Hubble function, depicted in the right panel of Fig.~\ref{fig3}, is a monotonically decreasing function of time. The time variations of both $V$ and $h$ are practically independent on the modifications of the numerical values of the parameter $\mu _0$, with a very slight influence manifesting itself only in the large time limit. On the other hand for large values of $\tau $, $h$ tends to a constant value, indicating that in the presence of the Maxwell-Chern Simons terms the dust fluid filled Bianchi type I Universe experiences a transition to an accelerating phase, ending in a de Sitter type regime. The time evolution of the vector field $B_0$, represented in the left lower panel of Fig.~\ref{fig3}, indicates a strong dependence on the numerical values of $\mu _0$. $B_0$ is a monotonically decreasing function of time, and it continues to decrease even during the accelerated expansion of the Bianchi type I geometry, and in the isotropic phase. The energy density $r$ of the dust matter, depicted in right lower panel of Fig.~\ref{fig3}, is a monotonically decreasing function of time, whose dynamics is essentially independent on the modifications of the numerical values of $\mu _0$. In the large time limit the matter energy density tends to zero, $\lim _{\tau \rightarrow \infty}r=0$, thus showing that the accelerated expansion of the dust Bianchi type I Universe in the presence of a Maxwell-Chern-Simons type field leads in its final stages to a vacuum Universe, in which the energy density of the ordinary matter gives a negligibly contribution to the total energy of the Universe.

The time variations of the deceleration parameter and of the anisotropy parameter of the dust Bianchi type I Universe in the presence of the Maxwell-Chern-Simons field are plotted in Fig.~\ref{fig4}.

The time variation of the deceleration parameter is presented in the left panel of Fig.~\ref{fig4}. For the chosen numerical values of the model parameters the expansion of the Bianchi type I Universe starts with values of $q$ of the order of 1.5,  $q\approx 1.5$ at $\tau =0$. The dust Bianchi I Universe is in a decelerating phase for small values of $\tau $, with $q>0$. However,  for $\tau =\tau _{cr}\approx 0.9$, the deceleration parameter reaches its borderline  value $q\approx 0$, and for $\tau >\tau _{cr}$ the Universe enters in an accelerating phase, with $q<0$. In the large time limit $q$ reaches values of the order of $q\approx -1$. The time evolution of $q$ is basically independent on the  numerical values of $\mu _0$. The time variation of the anisotropy parameter $A$, depicted in the right panel of Fig.~\ref{fig4}, does show a mild dependence on the numerical values of the parameter $\mu _0$. In the large time limit $A$ becomes zero, $A\rightarrow 0$. This situation corresponds to values of $q$ of the order of $q\approx -1$, which shows that the Universe is already isotropic before entering the de Sitter phase. The large time behavior of $A$, before the full isotropization period, shows a small dependence on the numerical values of $\mu _0$.

\section{Discussions and final remarks}

In this paper we have considered the effects of the topological Chern-Simons term on the  cosmological evolution of the Universe. The original Chern-Simons term consists of a constant vector field, which breaks the Lorentz invariance of the theory. In order to restore the Lorentz symmetry, we have promoted this constant vector to a dynamical vector field, with constant norm, through a condition which is imposed in the action by a Lagrange multiplier term. In this way we have obtained a bi-vector tensor theory of gravity. Another example of a bi-vector gravity theory was considered in \cite{WCGB} where the Weyl vector and the trace of torsion tensor are two vector fields coupled minimally to gravity. The norm of the vector field $A_\mu$ in our model could in principle be positive or negative or even zero. However, in order to respect the spatial isotropy of the Universe in the FRW space-time this constrained vector field should be time-like. The theory has a self-accelerating de Sitter solution. We have performed the cosmological perturbation analysis of the theory around this de Sitter background, and we have found that the theory has 2 gravitational wave modes, together with 4 vector modes, corresponding to the two vector fields of the theory. The vector modes of the theory interact with each other through the Chern-Simons interaction term. This interaction makes the vector degrees of freedom of the theory non trivial. Also there is no scalar mode in this theory, leaving 6 degrees of freedom  around the de Sitter background. It should be mentioned that the constrained vector field $A_\mu$ should have 3 dynamical degrees of freedom around the de Sitter background, since we have a self-interaction potential in the action. However, because of the Lagrange multiplier term, the helicity-0 mode will become non-dynamical in this case. This is similar to the case Einstein-aether theory where a constrained time-like vector field is added to the Einstein-Hilbert action. The main difference is that here we have a bi-vector tensor theory with a non-trivial Chern-Simons interaction term. The perturbation analysis of the theory then shows that the theory is stable and healthy around de Sitter background. Of course we have to perform Hamiltonian analysis in order to find the exact number of degrees of freedom of the theory.

In order to investigate the cosmological implications of the space-like constrained vector field one should consider an anisotropic space-time. In this paper, we have considered the dynamical behavior of a Bianchi type I space-time in the framework of the Maxwell-Chern-Simons gravity model. The Bianchi type I anisotropic space-time represents the simplest, and most natural, extension of the standard FRW metric, to which it reduces in the particular limit of equal directional scale factors. Bianchi type models can be considered as viable alternatives to the standard flat FRW cosmologies \cite{Bg}. The small observed deviations  from the exact isotropy and the anomalies in the Cosmic Microwave Background could be explained by the presence of an anisotropic expansion of the Universe. Bianchi type VIIh anisotropic cosmological models were considered in \cite{B7}, in a tentative to explain the large scale asymmetry observed in the Cosmic Microwave Background distribution. An in-depth comparison with the WMAP first-year data on large angular scales did show  that a chance alignment can be eliminated at a 3$\sigma $ level. On the other hand,  the recent Planck Collaboration results \cite{Planck} did show convincingly that the Bianchi type VIIh anisotropic cosmological model cannot fit the recent observational data obtained by the Planck satellite. However, one of the large angle anomalies of the Cosmic Microwave Background, the low quadrupole moment, indicates a great amount of power suppression at large scales. The presence of such an anomaly seems to indicate the presence of a Bianchi type I anisotropic geometry of the Universe. The extreme smallness of the quadrupole component of the Cosmic Microwave Background temperature distribution seems to suggest that for a homogeneous but anisotropic Universe the deviation from the isotropic flat FRW geometry must be small. Therefore such a deviation can be naturally explained by the existence of a background Bianchi type I geometry.

In the present paper we have found that in the framework of Maxwell-Chern-Simons  gravity theory the Bianchi type I homogeneous but anisotropic Universe presents a complex dynamics.  In our analysis we have assumed that the matter content of the Universe consists of a perfect barotropic cosmological fluid. In particular, for this type of matter source, the Bianchi type I models we have considered do always isotropize. The nature of the cosmological evolution strongly depends on the dimensionless model parameters $\lambda _1$ and $\lambda _2$, as well as on the adopted initial conditions for the matter energy density, Hubble function, and of the vector field itself. The transition to an isotropic phase of the Universe is associated with an accelerated, de Sitter type expansion, in which the considered cosmological models end in the large time limit. This type of behavior is independent on the nature of the cosmological fluid, and it does appear in both the cases of the stiff and dust fluids, respectively.

 Hence, the inclusion in the gravitational action of two vector fields, coupled via a Chern-Simons term, leads to the possibility of obtaining more general gravitational models, thus allowing for the possibility of a better physical description of both the very early and late evolution of our Universe.


\begin{thebibliography}{9}
\bibitem{acc-papers} A. G. Riess et al., Astron. J. {\bf 116}, 1009 (1998); S. Perlmutter et al., Astrophys. J. \textbf{517}, 565 (1999); R. A. Knop et al., Astrophys. J. {\bf 598}, 102 (2003); R. Amanullah et al., Astrophys. J. {\bf 716}, 712 (2010).
\bibitem{cos-const-prob} S. Weinberg, Rev. Mod. Phys. {\bf 61}, 1 (1989).
\bibitem{fRr} A. De Felice, S. Tsujikawa, Living Rev. Rel. {\bf 13}, 3 (2010); T. P. Sotiriou, V. Faraoni, Rev. Mod. Phys. {\bf 82}, 451 (2010).
\bibitem{fRRmunu} O. Bertolami, C. G. Boehmer, T. Harko, and F. S. N. Lobo, Phys. Rev. {\bf D 75}, 104016 (2007); T. Harko and F. S. N. Lobo, Eur. Phys. J. {\bf C 70}, 373 (2010); T. Harko, F. S. N. Lobo, S. Nojiri, S. D. Odintsov, Phys. Rev. {\bf D 84}, 024020 (2011); Z. Haghani, T. Harko, F. S. N. Lobo, H. R. Sepangi, S. Shahidi, Phys. Rev. {\bf D 88}, 044023 (2013); F. S. N. Lobo, T. Harko, arXiv:1211.0426v1 [gr-qc]; T. Harko, F. S. N. Lobo, Eur. Phys. J. {\bf C 70}, 373 (2010); S. D. Odintsov, D. Sáez-Gómez, Phys. Lett. {\bf B 725},  437 (2013); N. Katırcı and M. Kavuk, Eur. Phys. J. Plus {\bf 129}, 163 (2014).
\bibitem{massive} C. de Rham, G. Gabadadze, A. J. Tolley, Phys. Rev. Lett. {\bf 106}, 231101 (2011); C. de Rham, G. Gabadadze, Phys. Rev. {\bf D 82}, 044020 (2010); G. D'Amico, G. Gabadadze, L. Hui, D. Pirtskhalava, Phys. Rev. {\bf D 87}, 064037 (2013); C. de Rham, G. Gabadadze, Phys. Lett. {\bf B 693}, 334 (2010).
\bibitem{kessence} C. Armendariz-Picon, V. Mukhanov, Paul J. Steinhardt, Phys. Rev. {\bf D 63}, 103510 (2001); J. Yoo, Y. Watanabe, Int. J. Mod. Phys. {\bf D 21}, 1230002 (2012); S. Tsujikawa, arXiv:1004.1493 [astro-ph.CO]; E. J. Copeland, M. Sami, S. Tsujikawa, 	Int. J. Mod. Phys. {\bf D 15}, 1753 (2006).
\bibitem{quintessence} S. Tsujikawa, Class. Quantum Grav. {\bf 30}, 214003 (2013); J. Martin, Mod. Phys. Lett. {\bf A 23}, 1252 (2008).
\bibitem{inf} D. Baumann, arXiv:0907.5424 [hep-th]; I. P. Neupane, C. Scherer, JCAP {\bf 0805}, 009 (2008); N. Bose, A. S. Majumdar, Phys. Rev. {\bf D  79}, 103517 (2009).
\bibitem{DE} M. J. Mortonson, D. H. Weinberg, M. White, arXiv:1401.0046 [astro-ph.CO]; J. Frieman, M. Turner, D. Huterer, Ann. Rev. Astron. Astrophys. {\bf 46}, 385 (2008); L. García, J. Tejeiro, L. Castañeda, 	arXiv:1210.5259 [astro-ph.CO].
\bibitem{gali} A. Nicolis, R. Rattazzi, E. Trincherini, Phys. Rev. {\bf D 79}, 064036 (2009); C. Deffayet, G. Esposito-Farese, A. Vikman, Phys. Rev. {\bf D 79}, 084003 (2009);  G. W. Horndeski, Int. J. Theor. Phys. {\bf 10}, 363 (1974).
\bibitem{gali-cos} N. Chow, J. Khoury, Phys. Rev. {\bf D 80}, 024037 (2009); F. P Silva, K. Koyama, Phys. Rev. {\bf D 80}, 121301 (2009); A. De Felice, S. Tsujikawa, Phys. Rev. Lett. {\bf 105}, 111301 (2010); R. Gannouji, M. Sami, Phys. Rev. {\bf D 82}, 024011 (2010); D. F. Mota, M. Sandstad, T. Zlosnik, JHEP {\bf 1012}, 051 (2010); C. Burrage, C. de Rham, D. Seery and A. J. Tolley, JCAP {\bf 1101}, 014 (2011); A. De Felice, S. Tsujikawa, Phys. Rev. {\bf D 84}, 124029 (2011); C. Burrage, C. de Rham, L. Heisenberg, JCAP {\bf 1105}, 025 (2011).
\bibitem{emax} W. C. dos Santos, arXiv:1606.08527 [gr-qc].
\bibitem{max-proca} M. Dunajski, J. Gutowski, W. Sabra, P. Tod, Class. Quantum Grav. {\bf 28}, 025007 (2011); S. S. Yazadjiev, Bulg. J. Phys. {\bf 27}, 58 (2000); T. Maki, K. Shiraishi, Class. Quantum Grav. {\bf 10}, 2171  (1993); A. M. Ghezelbash, Phys. Rev. {\bf D 91}, 084003 (2015); C. Vuille, J. Ipser, J. Gallagher, arXiv:1406.0497 [gr-qc]; D. N. Vollick, arXiv:gr-qc/0601016v1; T. Dereli, M. Onder, J. Schray, R. W. Tucker, C. Wang, Class. Quantum Grav. {\bf 13}, L103 (1996); D. Klemm, M. Nozawa, M. Rabbiosi, Class. Quantum Grav. {\bf 32}, 205008 (2015); A. M. Ghezelbash and V. Kumar, arXiv:1704.01476 [gr-qc]; R. Tibrewala, Class. Quantum Grav. {\bf 29}, 235012 (2012); Y. Lim, arXiv:1702.05201 [gr-qc]; J. Bl\'{a}zquez-Salcedo, J. Kunz, F. Navarro-L\'{e}rida and E. Radu, Entropy {\bf 18}, 438 (2016);  Y. Ling, Z. Xian and Z. Zhou, Chinese Physics {\bf C 41}, 023104 (2017); C. Vuille, J. Ipser and J. Gallagher, Gen. Rel. Grav. {\bf 34}, 689 (2002); B. G\'{o}nzales, R. Linares, M. Maceda and O. S\'{a}nchez-Santos, arXiv:1409.3759 [hep-th]; Y. N. Obukhov and E. J. Vlachynsky, Ann. Phys. {\bf 8}, 497 (1999); H. Liu, H. Lu and C. N. Pope, J. High Energ. Phys. {\bf 2014}, 109 (2014); G. Li, Y. Zhang, L. Zhang, Z. Feng and X. Zu, 	Int. J. Theor. Phys. {\bf 54}, 1245 (2015); C. Herdeiro, E. Radu and H. Runarsson, Class. Quantum Grav. {\bf 33}, 154001 (2016); M. Minamitsuji, Phys. Rev. {\bf D 94}, 084039 (2016).
\bibitem{weyl-review} J. Attard, J. François, and S. Lazzarini, Phys. Rev. {\bf D 93}, 085032 (2016); J. Miritzis, J. Phys.: Conf. Ser. {\bf 8}, 131 (2005); C. Romero, J. B. Fonseca-Neto, M. L. Pucheu, Class. Quantum Grav. {\bf 29}, 155015 (2012); E. Scholz, arXiv:1111.3220 [math.HO].
\bibitem{many-weyl} A. T. Filippov, arXiv:0812.2616 [gr-qc]; E. Elizalde, G. Cognola and S. Zerbini, arXiv:1301.6269 [gr-qc]; M. Nishioka, Nuov. Cim. {\bf A 78}, 462 (1983); A. N. Makarenko and V. V. Obukhov, Russ. Phys. J. {\bf 41}, 1124 (1998); H. Lu, Y. Pang, C. N. Pope and J. Vazquez-Poritz, Phys. Rev. {\bf D 86}, 044011 (2012); Y. S. Myung, Phys. Lett. {\bf B 728}, 422 (2014); J. Peng, Eur. Phys. J. {\bf C 74}, 3156 (2014); M. Dunajski, J. Gutowski and W. Sabra, Class. Quantum Grav. {\bf 34}, 045009 (2017); D. Klemm, Class. Quantum Grav. {\bf 15}, 3195 (1998).
\bibitem{tomi} J. B. Jimenez and T. S. Koivisto, Class. Quantum Grav. {\bf 31}, 135002 (2014).
\bibitem{cos-weyl} J. B. Jimenez, L. Heisenberg and T. S. Koivisto, JCAP {\bf 1604}, 046 (2016).
\bibitem{nogo} C. Deffayet, A. E. Gumrukcuoglu, S. Mukohyama, Y. Wang, J. High Energ. Phys. {\bf 2014}, 82 (2014).
\bibitem{vec-gali} L. Heisenberg, JCAP {\bf 05}, 015 (2014); M. Hull, K. Koyama, G. Tasinato, Phys. Rev. {\bf D 93}, 064012 (2016).
\bibitem{cos-vec-gali} J. B. Jimenez, arXiv:1606.04361 [gr-qc]; A. De Felice, L. Heisenberg, R. Kase, S. Mukohyama, S. Tsujikawa, Y. Zhang, JCAP {\bf 1606}, 048 (2016).
\bibitem{gene-vec-gali} S. Nakamura, R. Kase, S. Tsujikawa, arXiv:1702.08610 [gr-qc]; J. B. Jimenez, L. Heisenberg, Phys. Lett. {\bf B 763}, 002 (2017); E. Allys, P. Peter, Phys.Rev. {\bf D 94}, 084041 (2016); L. Heisenberg, R. Kase, S. Tsujikawa, Phys. Lett. {\bf B 760}, 617 (2016).
\bibitem{BP} Z. Haghani, T. Harko, H. R. Sepangi, S. Shahidi, Eur. Phys. J. {\bf C 77}, 137 (2017).
\bibitem{cartan} F. W. Hehl, Gen. Rel. Grav. {\bf 4}, 333 (1973); F. Gronwald, F. W. Hehl, arXiv:gr-qc/9602013v1; Z. Haghani, T. Harko, H. R. Sepangi, S. Shahidi, Phys. Rev. {\bf D 88}, 044024 (2013); Z. Haghani, T. Harko, H. R. Sepangi, S. Shahidi, JCAP {\bf 10}, 061 (2012); D. Palle, J. Exp. Theor. Phys. {\bf 118}, 587 (2014); A. Trautman, arXiv:gr-qc/0606062.
\bibitem{WCGB} Z. Haghani, N. Khosravi, S. Shahidi, Class. Quantum Grav. {\bf 32}, 215016 (2015).
\bibitem{EA} T. Jacobson, D. Mattingly, Phys. Rev. {\bf D 64}, 024028 (2001); T. Jacobson, arXiv:0801.1547 [gr-qc].
\bibitem{horavarelation} T. Jacobson, Phys. Rev. {\bf D 81}, 101502 (2010); E. Barausse, T. Jacobson, T. P. Sotiriou, Phys. Rev. {\bf D 83}, 124043 (2011).
\bibitem{SEA} Z. Haghani, T. Harko, H. R. Sepangi, S. Shahidi, arXiv:1404.7689 [gr-qc]; T. Jacobson, A. J. Speranza, arXiv:1405.6351 [gr-qc]; D. Blas, O. Pujol\'{a}s, and S. Sibiryakov, J. High Energy Phys. {\bf 10}, 029 (2009).

 \bibitem{v1} C. Armendariz-Picon, J. Cosmol. Astropart. Phys. \textbf{07} (2004) 007.

\bibitem{v2} V. V. Kiselev, Class. Quantum Grav. \textbf{21}, 3323 (2004);
H. Wei and R.-G. Cai, Phys. Rev. \textbf{D 73}, 083002 (2006); T. S.
Koivisto and D. F. Mota, J. Cosmol. Astropartic. Phys. \textbf{0808}, 021
(2008); J. B. Jim\'{e}nez and A. L. Maroto, Phys. Rev. \textbf{D 78}%
, 063005 (2008); J. Beltr\'{a}n Jim\'{e}nez, R. Lazkoz, and A. L. Maroto,
Phys. Rev. \textbf{D 80}, 023004 (2009); V. V. Lasukov, Russian Physics
Journal \textbf{53} 296 (2010); E. Carlesi, A. Knebe, G. Yepes, S.
Gottloeber, J. Beltr\'{a}n Jim\'{e}nez, and A. L. Maroto, Monthly Not. Royal
Astron. Soc. \textbf{418}, 2715 (2011); E. Carlesi, A. Knebe, G. Yepes, S.
Gottloeber, J. Beltr\'{a}n Jim\'{e}nez, Antonio L. Maroto, Monthly Not.
Royal Astron. Soc. \textbf{424}, 699 (2012); N. Br\'et{o}n, ``\textit{Vector Fields Resembling Dark Energy}", in C. Moreno González et al. (eds.), Accelerated Cosmic Expansion,
Astrophysics and Space Science Proceedings Vol. {\bf 38} (Springer International Publishing,  New York, Dordrecht, London 2014), p. 61.

\bibitem{v3} C. G. B\"ohmer and T. Harko, Eur. Phys. J. \textbf{C 50}, 423
(2007).

\bibitem{SupracondDE} S.-D. Liang and T. Harko,
Phys. Rev.  \textbf{ D 91}, 085042 (2015); Z. Keresztes, L. A. Gergely, T. Harko, and S.-D. Liang, Phys. Rev. {\bf D 92}, 123503 (2015).


\bibitem{undota} S. M. Carroll, G. B. Field and R. Jackiw, Phys. Rev. {\bf D 41}, 1231 (1990); E. Kant, F.R. Klinkhamer, Nucl. Phys. {\bf B 731}, 125 (2005).

\bibitem{2p1} R. Jackiw and S. Templeton, Phys. Rev. {\bf D 23}, 2291 (1981); S. Deser, R. Jackiw and S. Templeton, Phys. Rev. Lett. {\bf 48}, 975 (1982).
\bibitem{qed} A. P. Balachandran, L. Chandar, E. Ercolessi, T. R. Govindarajan and R. Shankar, Int. J. Mod. Phys. {\bf A 09}, 3417 (1994); P. Tan, B. Tekin and Y. Hosotani, Nucl. Phys. {\bf B 502}, 483 (1997); S. S. Madrigal, C. P. Hofmann and A. Raya, J. Phys.: Conf. Ser. {\bf 287}, 012028 (2011).
\bibitem{epjc} R. Casana, M. M. Ferreira, E. da Hora and A. B. F. Neves, Eur. Phys. J. {\bf C 74}, 3064 (2014).
\bibitem{bound} A. Blasi, N. Maggiore, N. Magnoli and S. Storace, Class. Quantum Grav. {\bf 27}, 165018 (2010).


 \bibitem{CS1} A. Lue, L. M. Wang and M. Kamionkowski, Phys. Rev. Lett. {\bf 83}, 1506 (1999).
  \bibitem{CS2} R. Jackiw and S.-Y. Pi, Phys. Rev. {\bf D 68}, 104012 (2003).
  \bibitem{CS3}  S. Alexander and N. Yunes,   Phys. Rept. {\bf 480}, 1 (2009).
  \bibitem{CS4}  B. A. Campbell, N. Kaloper, R. Madden and K. A. Olive, Nucl. Phys. {\bf B 399}, 137 (1993); S. H. S. Alexander and S. J. J. Gates, JCAP {\bf 0606}, 018 (2006).
      \bibitem{CS5} T. Harko, Z. Kov\'{a}cs, and F. S. N. Lobo, Class. Quantum Grav. {\bf 27}, 105010 (2010).

\bibitem{quant} D. Colladay, V. A. Kostelecký, Phys. Rev. {\bf D 58}, 116002 (1998); C. Adam, F. R. Klinkhamer, Nucl. Phys. {\bf B 607}, 247 (2001).

\bibitem{An} C.-M. Chen, T. Harko, and M. K. Mak, Phys. Rev. {\bf D 62}, 124016 (2000); T. Harko and M. K. Mak,  Class. Quantum Grav. {\bf 21},  1489 (2004); T. Harko, F. S. N. Lobo, and M. K. Mak, Galaxies {\bf 2}, 496 (2014).
\bibitem{Zel} S. L. Shapiro and S. A. Teukolsky, ``\textit{Black Holes, White Dwarfs and Neutron Stars}", John Wiley \& Sons, New York, NY, USA, 1983

\bibitem{Bg} R. Moriconi, G. Montani, and S. Capozziello, Phys. Rev. {\bf D 94}, 023519, (2016);
 M. Sharif and A. Siddiqa, Phys. Lett. {\bf A 381},  838 (2017); E. Bittencourt, L. G. Gomes, and R. Klippert, Class. Quantum Grav. {\bf 34},  045010 (2017); A. Yu. Kamenshchik, E. O. Pozdeeva, S. Yu. Vernov, A. Tronconi, and G. Venturi, Phys. Rev. {\bf D 95}, 083503 (2017).

\bibitem{B7}  T. R. Jaffe, A. J. Banday, H. K. Eriksen, K. M. Gorski, and F. K. Hansen, Astrophys. J.  {\bf 629}, L1 (2005).

\bibitem{Planck} P. A. R. Ade et al., Astron. Astrophys. {\bf 594}, A18 (2016).
\end{thebibliography}
\end{document}